\documentclass[preprint,12pt,number,sort&compress]{elsarticle}
\pdfoutput=1
\usepackage{graphicx}
\usepackage{latexsym,amssymb,amsmath}
\usepackage{bm}
\usepackage[caption=false]{subfig}
\usepackage{xcolor}
\usepackage{mathrsfs}
\usepackage{hyperref}
\begin{document}
\begin{frontmatter}
\title{Effect of Dependent Scattering on Light Absorption in Highly Scattering Random Media}
\author{B. X. ~Wang}
\author{C. Y. ~Zhao\corref{cor1}}
\ead{Changying.zhao@sjtu.edu.cn}
%\thanks{}
\address{Institute of Engineering Thermophysics, Shanghai Jiao Tong University, Shanghai, 200240, P. R. China}
\cortext[cor1]{Corresponding author}

\begin{abstract}
The approximate nature of radiative transfer equation (RTE) leads to a bunch of considerations on the effect of “dependent scattering” in random media, especially particulate media composed of discrete scatterers, in the last a few decades, which usually indicates those deviations RTE (combined with ISA) lead to from experimental and exact numerical results due to electromagnetic wave interference. Here we theoretically and numerically demonstrate the effect of dependent scattering on absorption in disordered media consisting of highly scattering scatterers. By making comparison between the independent scattering approximation-radiative transfer equation (ISA-RTE) and the full-wave coupled dipole method (CDM), we find that deviations between the two methods increase as scatterer density in the media increases. The discrepancy also grows with optical thickness. To quantitatively take dependent scattering effect into account, we develop a theoretical model using quasi-crystalline approximation (QCA) to derive dependent-scattering corrected radiative properties, based on the path-integral diagrammatic technique in multiple scattering theory. The model results in a more reasonable agreement with numerical simulations. The present work has profound implications for the coherent scattering physics in random media with absorption, correctly modeling light absorptance in random media and interpreting the experimental observations in various applications for random media such as solar energy concentration, micro/nanofluids, structural color generation, etc.
\end{abstract}
\end{frontmatter}

\section{Introduction}\label{intro}
Light propagation in complex and random media is a very general problem in many disciplines of science and engineering, such as radiative heat transfer \cite{tien1987thermal,modest2013radiative}, optics and photonics \cite{wiersma2013disordered,Rotter2017}, astrophysics and remote sensing \cite{tsangRS2000,mishchenko2006multiple}, soft mater physics and chemistry \cite{xiaoSciAdv2017}, biomedical engineering \cite{horstmeyer2015guidestar} and so on. Generally, in such media, light is scattered and absorbed in a very complicated way, which not only depends on the features of incident light (frequency and polarization state, etc.), but also heavily depends on the properties of media, including the permittivity and permeability as well as their frequency and spatial dispersion (if any) of the composing materials, the morphology (size, shape and topology) of inclusions, and their time- and temperature-dependency (if any). Conventional parameters describing light propagation in random media, including scattering coefficient $\mu_s$, absorption coefficients $\mu_a$ and scattering phase function $P(\mathbf{\Omega}',\mathbf{\Omega})$, are defined in the framework of radiative transfer equation (RTE) \cite{VanRossum1998,tsang2004scattering}. The RTE, although derived phenomenologically in its initial stage, is actually an approximate form of Bethe-Salpter equation for classical photons \cite{lagendijk1996resonant,VanRossum1998,tsang2004scattering,sheng2006introduction}. The latter is an rigorous equation accounting for all interference phenomena for the transport of electromagnetic field correlation function $\langle\mathbf{E}\mathbf{E}^*\rangle$ in random media, which is originally taken from quantum field theory and 
is exactly equivalent to Maxwell equations for electromagnetic waves. 

Particularly, for discrete random media composed of a disordered distribution of scatterers, the RTE is valid when the following two conditions are simultaneously satisfied: (1) the scatterers are far apart from each other and each scatterer scatters light as if no other scatters exist. This is called independent scattering approximation (ISA); (2) the interference between each multiple scattering trajectory and its time-reversal counterpart is neglected \cite{VanRossum1998,mishchenko2006multiple,sheng2006introduction} . The latter condition is called ladder approximation because it results in the Feynmann diagrammatic representation of field correlation function resembling a series of ladders \cite{VanRossum1998,mishchenko2006multiple,sheng2006introduction}. The approximate nature of RTE leads to a bunch of considerations on the effect of ``dependent scattering'' in the last a few decades, which usually indicates those deviations RTE (combined with ISA) lead to from experimental and exact numerical results due to electromagnetic wave interference \cite{tien1987thermal,kumar1990dependent,leeJTHT1992,ivezicIJHMT1996,durantJOSAA2007,nguyenOE2013,wangIJHMT2015,maJQSRT2017}. Note herein ``dependent scattering effect'' is termed as a generalization for those interference effects that are not possible to explain under independent scattering approximation (ISA) \cite{yamadaJHT1986,aernoutsOE2014}. This is a very broad definition and different from the meaning of van Tiggelen et al.'s \cite{Vantiggelen1990JPCM}, where they classified the multiple scattering trajectories visiting the same particle more than once and resulting in a closed loop, or ``recurrent scattering'', as the dependent scattering mechanism. 

There are several dependent scattering mechanisms recognized both theoretically and experimentally already, including recurrent scattering \cite{Vantiggelen1990JPCM, Cherroret2016}, coherent backscattering (also called weak localization) \cite{mishchenko2006multiple} and the well-known Anderson (strong) localization of light \cite{wiersma1997localization}. Note these mechanisms are not entirely independent concepts with each other. Typically, the dependent scattering mechanisms are treated in a renormalized way, i.e., by correcting conventional radiative parameters into effective parameters to include dependent scattering effects and retaining the form of RTE and diffusion equation (an approximate form of RTE) to solve the transport problem \cite{tsang2004scattering}.

Most researches up to now focus on the role of dependent scattering mechanism in scattering properties of random media, because its impact on scattering properties is more prominent and can result in attractive phenomena as mentioned above. In terms of its role in absorption, very few studies, to the best of our knowledge, exist \cite{kumar1990dependent,ma1990enhanced,prasherJAP2007,weiAO2012}. Actually, when the particle density is small (typically volume fraction $f_v<0.05$) and absorption coefficient is much larger than the scattering coefficient, i.e., $\mu_a\gg\mu_s$, this ignorance gives rise to no substantial discrepancies because multiple and dependent scattering is weak. However, when particle density continues to increase or $\mu_s$ is comparable with or much larger than $\mu_a$, a careful consideration of dependent scattering effect on total absorption is necessary because the interparticle interference of scattering waves may lead to a redistribution in particle absorption. This issue is becoming important as the recent growing interests in nanofluids, as well as other nanoparticle-based solar absorbers, which usually utilize plasmonic resonances of metallic particles to enhance solar absorption \cite{taylorJAP2013,saidICHMT2013,xuanRSCA2014,hoganNL2014,liuNanoscale2017,gaoJAP2017}, finding their applications in concentrating solar power and direct-steam generation. In this situation, the scattering coefficient might be comparable with the absorption coefficient for plasmonic particles with diameter approaching 100nm. Actually this feature is exploited by some authors because multiple scattering of light can enhance the light path length in the medium and thus improve the absorption efficiency \cite{hoganNL2014}. Note the ``multiple scattering'' in this context is under the framework of RTE, meaning that light \textit{intensity} is multiply scattered many times, without any considerations on wave aspect of the coherent scattering problem. Here throughout the rest of this paper, we term ``multiple scattering'' as the multiple scattering of \textit{electromagnetic waves} implicitly. In fact in this circumstance, the strong scattering strength can also lead to a strong modification of the local electromagnetic field to be different with external incident field, as a manifestation of dependent scattering mechanism \cite{kumar1990dependent,ma1990enhanced}. Wei et al.\cite{weiAO2012} recently investigated the effect of dependent scattering on absorption coefficient for very dense nanofluids ($f_v$ up to 0.74) containing very small metallic particles (radius $a=15\mathrm{nm}$) using several different dependent-scattering models as well as developed a modified quasicrystalline approximation (QCA) model. However, their model, as well as the model proposed by Prasher et al. \cite{prasherJAP2007} can only treat the particle absorption is much larger than its scattering, which is only valid for very small particles.

Another active field is optofluidics, a combination of micro/nano fluidics and photonics, which utilizes the easy reconfigurability and compactness of nanofluids through flow rate, viscosity, and so on, to obtain the desirable optical properties for integrated photonics applications, like lasers, sensors etc., as well as photocatalysis, solar thermochemistry and solar desalination applications. One of the most popular optofluidic systems is nanoparticle colloidal system controlled by microfluid chips, which shows the necessity of understanding the interplay between multiple scattering and absorption \cite{psaltisNature2006,ericksonNPhoton2011}. In some researches on structural color based on disordered photonic structures, predictions based on ISA can poorly interpret the experimental results, partly due to the extremely-high absorption predicted by ISA for the short-range ordered, densely packed nanostructures, which may smear out the reflectance peak indeed observed by the experiments \cite{xiaoSciAdv2017}. An in-depth physical interpretation for the scattering and absorption mechanism will also be helpful for these applications. 

Here we perform a rigorous study on the effect of dependent scattering on absorption for highly scattering particles, attempting to explore the role of dependent scattering. Deep understanding on the coherent phenomena will be given by making comparison between the (incoherent) radiative transfer model and coherent coupled dipole model. We also develop a theoretical model using quasi-crystalline approximation (QCA) to obtain dependent-scattering corrected radiative properties, based on the path-integral diagrammatic technique in multiple scattering theory. This model results in a more reasonable agreement with numerical simulations. The present study can provide physical insights on the applications of nanofluids in solar energy concentration, vapor generation using localized-light-induced-heat and heat transfer enhancement \cite{zielinskiNL2016}. Our results also have implications in the dipole-dipole interaction effect on absorption in other highly scattering random media.
\section{The Coupled Dipole Method}

We consider a slab geometry containing randomly distributed metallic particles in air as the investigated disordered medium with side lengths of $L_x$ and $L_y$, and a thickness of $L_z$. The incident light propagates along the direction $z$. Here the treatment of matrix as air doesn't lose the essential physics of multiple and dependent scattering, and can be easily extended to any kind of realistic matrix, such as gelatin, water, etc. We further assume that the scatterers are not so small that $a$ is larger than a few nanometers, which means that the bulk permittivity can be used to describe the small particles and nonlocal (or quantum) effect in permittivity can also be neglected. Moreover, we are working at the long wavelength limit, where the wavelength $\lambda$ is much larger than the small particle. In this way, the EM responses of individual dipoles are described by the dipole polarizability $\alpha$, which can be expressed by the first order electric Mie coefficient as \cite{bohrenandhuffman}:
\begin{equation}\label{alphamie}
\alpha=\frac{6\pi i}{k^3}\frac{m^2j_1(mx)[xj_1(x)]'-j_1(x)[mxj_1(mx)]'}{m^2j_1(mx)[xh_1(x)]'-h_1(x)[mxj_1(mx)]'}
\end{equation}
where $k=2\pi/\lambda$ is the wavenumber in vacuum, $x=ka=2\pi a/\lambda$ is the size parameter, $a$ is the radius of spherical particle, and $m=\sqrt{\varepsilon}$ is the complex refractive index of metal. $j_1$  and $h_1$ are first-order spherical Bessel functions and Hankel functions, respectively. 

\begin{figure}
	\includegraphics[width=\linewidth]{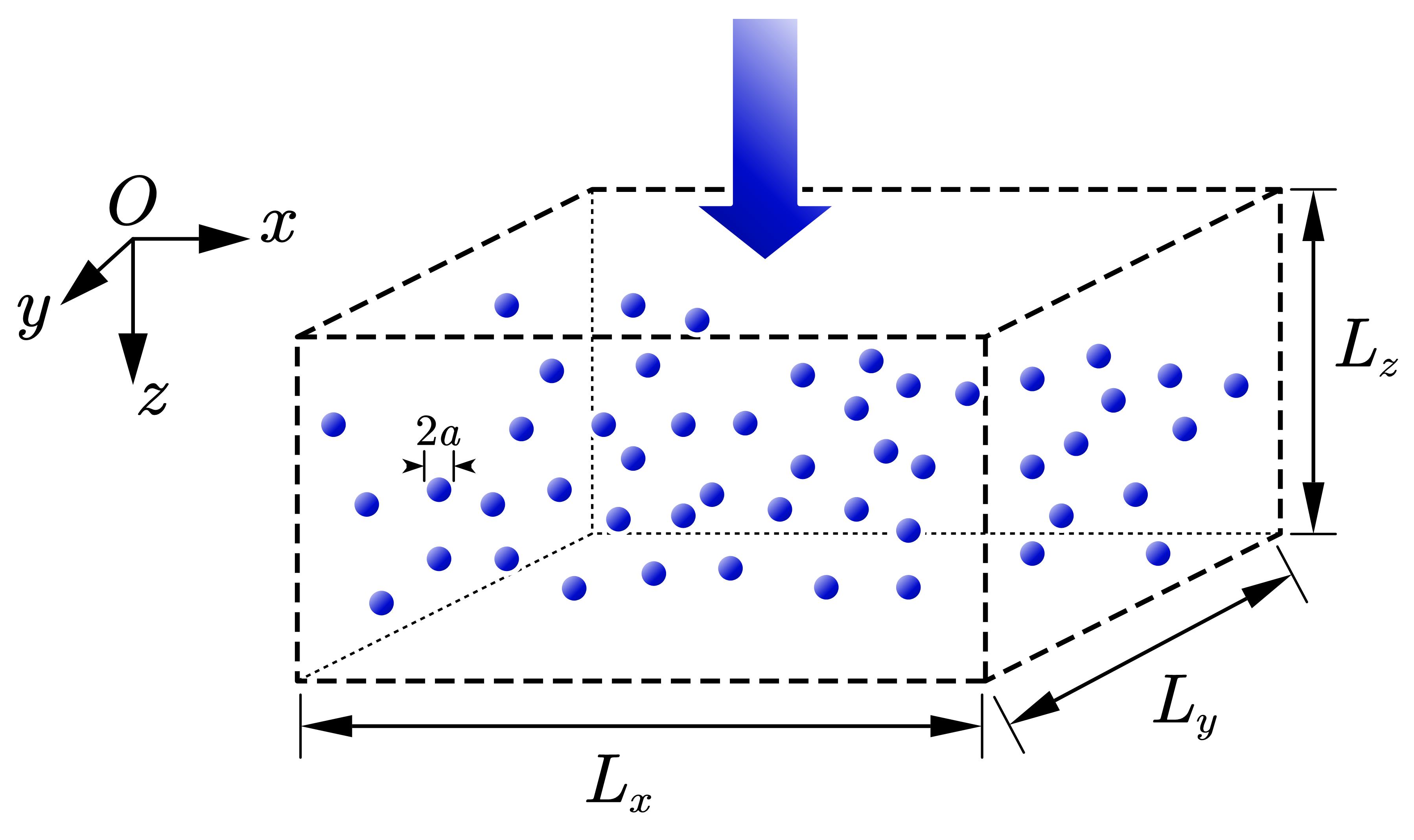}
	\caption{A random medium with a slab geometry consisting of randomly distributed spherical particles. Incident radiation is denoted by the large, solid arrow.}\label{system_config}
\end{figure}

Since the EM response of an individual dipole is described by the dipole polarizability $\alpha$, the coupled dipole method (CDM), as an exact solution for dipole scatterers distributed in the investigated random medium, is then used to calculate the exact total absorption of all particles. In vacuum or any homogeneous, isotropic host medium, the CDM has the following form \cite{markel1993}: 
\begin{equation}\label{coupled-dipole}
\mathbf{p}_j=\alpha\left[\mathbf{E}_{inc}(\mathbf{r}_j)+k^2\sum_{i=1,i\neq j}^{N}\mathbf{G}_0(\mathbf{r}_j,\mathbf{r}_i)\mathbf{p}_i\right]
\end{equation}
where $\mathbf{E}_{inc}(\mathbf{r}_j)$ is the incident field impinging on the $j$-th particle. Here we model the incident light as a plane wave leading to $\mathbf{E}_{inc}(\mathbf{r}_j)=\mathbf{E}_0\exp{(i\mathbf{k}\cdot\mathbf{r}_j)}$ with $\mathbf{k}=k\hat{\mathbf{z}}$. $\mathbf{p}_i$ is the excited dipole moment of $i$-th particle. $\mathbf{G}_{0}(\omega,\mathbf{r}_j,\mathbf{r}_i)$  is the free-space dyadic Green's function and describes the propagation of scattering field of $j$-th dipole to $i$-th dipole as \cite{markel1993,Cherroret2016} 
\begin{equation}
\begin{split}
\mathbf{G}_{0}(\mathbf{r}_j,\mathbf{r}_i)&=\frac{\exp{(ikr)}}{4\pi r}\left(\frac{i}{kr}-\frac{1}{k^2r^2}+1\right)\mathbf{I}\\
&+\frac{\exp{(ikr)}}{4\pi r}\left(-\frac{3i}{kr}+\frac{3}{k^2r^2}-1\right)\mathbf{\hat{r}}\mathbf{\hat{r}}+\frac{\delta({\mathbf{r}})}{3k^2}
\end{split}
\end{equation}
where the Dirac function $\delta(\mathbf{r})$ is responsible for the so-called local field in the scatterers \cite{Cherroret2016}. $\mathbf{I}$ is identity matrix and $\hat{\mathbf{r}}$ is the unit vector of $\mathbf{r}=\mathbf{r}_j-\mathbf{r}_i$. After calculating the EM responses of all scatterers based on above multiple scattering equations, we find the total scattering field of the random cluster of particles at an arbitrary position $\mathbf{r}\neq\mathbf{r}_j$ where $\mathbf{r}_j$ denotes the position of scatterers, is computed as
\begin{equation}
\mathbf{E}_s(\mathbf{r})=k^2\sum_{i=1}^{N}\mathbf{G}_0(\mathbf{r},\mathbf{r}_i)\mathbf{p}_i
\end{equation}
The total absorption cross section is given by
\begin{equation}
C_{abs}=k\sum_{j=1}^N\mathrm{Im}(\mathbf{p}_j\cdot\mathbf{E}_{exc,j}^*-\frac{k^3}{6\pi}|\mathbf{p}_j|^2)
\end{equation}
where $\mathbf{E}_{exc,j}$ is the exciting field imping on $j$-th particle and given by $\mathbf{E}_{exc,j}=\mathbf{p}_j/\alpha$. 
%The total extinction cross section is given by
%\begin{equation}
%C_{ext}=k\sum_{j=1}^N\mathrm{Im}\left[\mathbf{p}_j\cdot\mathbf{E}_{inc}^{*}(\mathbf{r}_j)\right]
%\end{equation}
Therefore, the total absorption for a slab geometry with a cross section $S$, whose normal vector is parallel to the propagation direction of incident light, is given by $A=C_{abs}/S$ when a plane wave illumination is applied. Here in this study, we regard CDM as an exact, standard method for solving the absorptance of the random media containing dipolar particles, and examine the validity of theoretical models by comparing their results with it.
\section{Independent Scattering Approximation and Radiative transfer equation}\label{RTE}
For a comparison with exact full-wave model for point scatterers, we present an incoherent Monte Carlo simulation based on radiative transfer equation. The conventional RTE relies on the radiative properties defined from independent scattering approximation (ISA) as aforementioned in Sec.\ref{intro}. Under this approximation, the extinction, scattering and absorption coefficients for randomly distributed dipolar particle systems are
\begin{equation}
\mu_{e}=n_0k\mathrm{Im}(\alpha)
\end{equation}
\begin{equation}
\mu_{s}=n_0\frac{k^4|\alpha|^2}{6\pi}
\end{equation}
\begin{equation}
\mu_{a}=n_0\left[k\mathrm{Im}(\alpha)-\frac{k^4|\alpha|^2}{6\pi}\right]
\end{equation}
where $n_0=N/V$ is the number density of the nanoparticles.
Thus using these parameters and radiative transfer equation (RTE), we are able to determine the total absorption of plasmonic systems without considering dependent scattering effects. Since RTE is derived under independent scattering approximation (ISA) for the field and ladder approximation for the intensity, which neglect all the interference effects, or dependent scattering effects, it is typically valid for dilute, uncorrelated media. Here we will compare the results from RTE to the numerically exact results to obtain the role of dependent scattering on absorption for a slab geometry. The calculated radiative properties, including scattering coefficients, absorption coefficients and phase functions, are substituted into the radiative transfer equation (RTE) to solve the hemisphere reflectance and transmittance of random media slabs shown in Fig.\ref{system_config}. The RTE for unpolarized light can be expressed as \cite{tsang2000scattering1}:
\begin{equation}
\frac{dI_\lambda}{ds}=\mu_aI_{b\lambda}-\mu_eI_\lambda+\frac{\mu_s}{4\pi}\int_{4\pi}I_\lambda(\Omega_i)P(\Omega,\Omega_i)d\Omega_i
\end{equation}
where  $I_\lambda$ is the spectral intensity,  $I_{b\lambda}$ is the spectral intensity of blackbody, $s$ is the transport path length and $P(\Omega,\Omega_i)$ is the phase function representing the scattering probability from direction $\Omega_i$  to direction $\Omega$. In the current study, the thermal radiation emission term $\mu_aI_{b\lambda}$ can omitted, since the reflectance and transmittance measurement is carried out in room temperature with sample emission far smaller than the incident light energy.

The random medium boundary here is treated as specular surface. Since the introduction of scatterers always changes the effective refractive index of the random media, when calculating specular reflectivity using Fresnel equations at the geometry boundary, the refractive index will not be 1 but the effective index of the random medium. The real part $n_{eff}$ of the effective complex refractive index $m_{eff}=\sqrt{\varepsilon_{eff}}$  is calculated under Maxwell-Garnett formula. MG is, to some extent, is adequate in determining the real part of refractive index, even in some photonic crystals and highly scattering media \cite{schuurmansScience1999}. The boundary effects are important in simulation, especially for highly scattering samples. The scattering phase function for unpolarized radiation is of the Rayleigh type \cite{bohrenandhuffman}
\begin{equation}\label{rayleigh_pf}
P(\Omega,\Omega_i)=\frac{3}{4}(1+\cos^2{\theta})
\end{equation}
where the scattering angle $\theta$ is the angle between incident direction solid angle $\Omega_i$ and scattering direction $\Omega$.

The Monte Carlo (MC) method, as a widely used technique to provide simple but excellent interpretation of very complex transfer problem for arbitrary geometries, is implemented here to trace the reflecting, scattering and absorption processes of a large amount of light energy bundles to solve the RTE. The total absorptance of the medium then can be  determined by simply adding up the absorbed weight of energy bundles inside the medium. A popular Monte Carlo code developed by Wang et al. is used in this study \cite{wang1995mcml} , where some modifications to consider specific phase function based on rejection sampling method are implemented \cite{wangIJHMT2015}. The details of Monte Carlo method can refer to Ref.\cite{wang1995mcml}.

\section{Dependent Scattering Model}
\subsection{General theory of multiple scattering}\label{general_theory}
When a large number of identical dipolar particles are randomly packed and constitute a disordered medium, it is of great interest to investigate whether the dependent scattering effect can lead to a modification of absorption behavior which ISA-RTE cannot predict. There are various models considering the interference effects in disordered systems containing randomly distributed small particles such as colloidal media, nanofluid and suspensions, including the well-known Maxwell-Garnett approximation (MGA) derived from Lorentz-Lorenz relation (LLR), i.e., the local field correction, the quasi-crystalline approximation (QCA) \cite{laxPR1952}, effective medium approximation (EMA) of Roth, Davis and Schwartz's \cite{rothPRB1974,davisPRB1985}, Persson and Liebsch's lattice-gas coherent potential approximation (LG-CPA) \cite{liebschJPC1983,liebschPRB1984}, and other  models taking nonlocal, correlation and multiplole effects into account were also developed in Refs. \cite{Felderhof1983,torquatoJCP1984,barreraPRB1988,barreraPRB1989,Claro1991,vasilevskiyPRB1996,felderhofPRE1998}. 

However, few models handled with the scattering coefficient induced by random fluctuations of particle density in disordered media, partly because their studies focused on very small particles in which scattering is much smaller than absorption, especially for those metallic particles with a radius $\sim5\mathrm{nm}$. It now comes into interest that for larger metallic particles, scattering cross section is usually comparable with or even larger than absorption cross section and the dependent scattering effect on absorption is appreciable. Here in this section, we will give analytical expressions of the multiple scattering problem for a random medium containing dipolar particles based on quasicrystalline approximation (QCA) to consider the dependent scattering effect on radiative properties of the present random medium \cite{laxPR1952,tsang2004scattering}.

Before proceeding to the theoretical derivation of radiative properties, let us give a brief overview of the multiple scattering theory, and demonstrate how this theoretical framework leads to the quasicrystalline approximation we will use to predict transport properties throughout the rest of the paper \cite{laxRMP1951,laxPR1952,lagendijk1996resonant,VanRossum1998, tsang2004scattering,sheng2006introduction}. In the general case of an infinite nonmagnetic three-dimensional (3D) medium, the spatial distribution of permittivity $\varepsilon(\mathbf{r})$ is inhomogeneous and can be generally described as $\varepsilon(\mathbf{r})=1+\delta \varepsilon(\mathbf{r})$, where $\delta \varepsilon(\mathbf{r})$ is the fluctuational part of the permittivity due to random morphology of the inhomogeneous medium, and electromagnetic wave propagation in such media is described by vector Helmholtz equation \cite{lagendijk1996resonant,tsang2004scattering}: 
\begin{equation}
\nabla \times \nabla \times \mathbf{E}(\mathbf{r})-k^{2}\varepsilon(\mathbf{r}) \mathbf{E}(\mathbf{r})=0
\end{equation}

Let $k^2=\omega^2/c_0^2$  be the wavenumber in the background medium and $V(\mathbf{r})=k^2\delta \varepsilon(\mathbf{r})=\omega^2\delta \varepsilon(\mathbf{r})/c_0^2$ be disordered potential inducing electromagnetic scattering, where $c_0$ is the speed of light in the background medium. Then we have an alternative form of vector Helmholtz equation convenient for random media problems,
\begin{equation}
\nabla \times \nabla \times \mathbf{E}(\mathbf{r})-k^{2} \mathbf{E}(\mathbf{r})=V(\mathbf{r})\mathbf{E}(\mathbf{r})
\end{equation}

To solve the equation, we introduce the dyadic Green's function for this random medium which satisfies
\begin{equation}
\nabla \times \nabla \times \mathbf{G}(\mathbf{r},\mathbf{r}')-k^{2} \mathbf{G}(\mathbf{r},\mathbf{r}')=V(\mathbf{r})\mathbf{G}(\mathbf{r},\mathbf{r}')+\mathbf{I}\delta(\mathbf{r},\mathbf{r}')
\end{equation}

In the meanwhile, the Green's function in the homogeneous background medium is 
\begin{equation}
\nabla \times \nabla \times \mathbf{G}_0(\mathbf{r},\mathbf{r}')-k^{2} \mathbf{G}_0(\mathbf{r},\mathbf{r}')=\mathbf{I}\delta(\mathbf{r},\mathbf{r}')
\end{equation}
Taking the Fourier transform from $\mathbf{r}$ and $\mathbf{r}'$ to its reciprocal space vector $\mathbf{p}$ and $\mathbf{p}'$ and let $V(\mathbf{r},\mathbf{r}')=k^2\delta \epsilon(\mathbf{r})\delta(\mathbf{r}-\mathbf{r}')$ using the Dirac delta function, we can write down the solution for dyadic Green's function in the disordered  media as
\begin{equation}\label{lp_eq1}
\mathbf{G}(\mathbf{p},\mathbf{p}')=\mathbf{G}_0(\mathbf{p},\mathbf{p}')+\mathbf{G}_0(\mathbf{p},\mathbf{p}')\mathbf{V}(\mathbf{p},\mathbf{p}')\mathbf{G}(\mathbf{p},\mathbf{p}')
\end{equation} 
which is known as the Lippman-Schwinger equation \cite{VanRossum1998,mishchenko2006multiple,sheng2006introduction}. Introducing the $T$-matrix $\mathbf{T}$, Eq.(\ref{lp_eq1}) is transformed into the following form
\begin{equation}\label{lp_eq2}
\mathbf{G}(\mathbf{p},\mathbf{p}')=\mathbf{G}_0(\mathbf{p},\mathbf{p}')+\mathbf{G}_0(\mathbf{p},\mathbf{p}')\mathbf{T}(\mathbf{p},\mathbf{p}')\mathbf{G}_0(\mathbf{p},\mathbf{p}')
\end{equation} 
It is easily shown that 
\begin{equation}
\mathbf{T}(\mathbf{p},\mathbf{p}')=[\mathbf{I}-\mathbf{V}(\mathbf{p},\mathbf{p}')\mathbf{G}_0(\mathbf{p},\mathbf{p}')]^{-1}\mathbf{V}(\mathbf{p},\mathbf{p}')
\end{equation}
If the medium only contains only one discrete scatterer, $\mathbf{T}(\mathbf{p},\mathbf{p}')$ is then known as the $T$-matrix for the single scatterer. Obviously for a random medium composed of many scatterers, Eq.(\ref{lp_eq1}) still applies. However, if each scatterer can be described by its own $T$-matrix, it is more convenient to transform Eq.(\ref{lp_eq1}) into the form only involving the $T$-matrices of the particles, rather than the ``scattering potential'' $\mathbf{V}$. In this manner, the Lippman-Schwinger equation for a medium consisting of $N$ discrete scatterers is rewritten as
\begin{equation}\label{lp_eq3}
\mathbf{G}(\mathbf{p},\mathbf{p}')=\mathbf{G}_0(\mathbf{p},\mathbf{p}')+\mathbf{G}_0(\mathbf{p},\mathbf{p}')\sum_{j=1}^N\mathbf{T}_j(\mathbf{p},\mathbf{p}')\mathbf{G}_0(\mathbf{p},\mathbf{p}')
\end{equation} 
where $\mathbf{T}_j(\mathbf{p},\mathbf{p}')=[\mathbf{I}-\mathbf{V}_j(\mathbf{p},\mathbf{p}')\mathbf{G}_0(\mathbf{p},\mathbf{p}')]^{-1}\mathbf{V}_j(\mathbf{p},\mathbf{p}')$ is the $T$-matrix of $j$-th scatterer. This equation is also known as Foldy-Lax equation for multiple scattering of classical waves \cite{foldyPR1945,laxRMP1951,mishchenko2006multiple}. To obtain statistically meaningful description of a random medium, it is necessary to take ensemble average of the full system to eliminate the impact of specific configuration. When taking ensemble average of Eq.(\ref{lp_eq3}), we obtain
\begin{equation}\label{lp_eq4}
\langle\mathbf{G}(\mathbf{p},\mathbf{p}')\rangle=\mathbf{G}_0(\mathbf{p},\mathbf{p}')+\mathbf{G}_0(\mathbf{p},\mathbf{p}')\langle\mathbf{T}(\mathbf{p},\mathbf{p}')\rangle\mathbf{G}_0(\mathbf{p},\mathbf{p}')
\end{equation} 
where $\langle\mathbf{G}(\mathbf{p},\mathbf{p}')\rangle$ denotes ensemble averaged amplitude Green's function, and $\langle\mathbf{T}(\mathbf{p},\mathbf{p}')\rangle$ is the ensemble averaged T-matrix of the full system, which is related to individual T-matrix of each scatterer in the manner of infinite multiple scattering process as
\begin{equation}\label{lp_eq5}
\begin{split}
\langle\mathbf{T}(\mathbf{p},\mathbf{p}')\rangle&=\langle\sum_{i=1}^N\mathbf{T}_j(\mathbf{p},\mathbf{p}')\rangle+\langle\sum_{i=1}^{N}\sum_{j\neq i}^{N}\mathbf{T}_j(\mathbf{p},\mathbf{p}_1)\mathbf{G}_0(\mathbf{p}_1,\mathbf{p}_2)\\&\times\mathbf{T}_j(\mathbf{p}_2,\mathbf{p}')\rangle+...
\end{split}
\end{equation} 
where the dummy variables $\mathbf{p}_1$ and $\mathbf{p}_2$ are integrated out and we don't write this procedure explicitly.  We then obtain the well-known Dyson equation for the coherent, or mean, component of the field \cite{lagendijk1996resonant,VanRossum1998,tsang2004scattering}
\begin{equation}
\langle\mathbf{G}(\mathbf{p},\mathbf{p}')\rangle=\mathbf{G}_0(\mathbf{p},\mathbf{p}')+\mathbf{G}_0(\mathbf{p},\mathbf{p}')\bm{\Sigma}(\mathbf{p},\mathbf{p}')\langle\mathbf{G}(\mathbf{p},\mathbf{p}')\rangle
\end{equation} 
where $\bm{\Sigma}(\mathbf{p},\mathbf{p}')$ is the so-called self-energy (or mass operator) containing all irreducible multiple scattering expansion terms in T-matrix $\langle\mathbf{T}(\mathbf{p},\mathbf{p}')\rangle$. The irreducible terms are those multiple scattering diagrams that cannot be divided without breaking the particle connections, including the same particle or particle correlations \cite{VanRossum1998}. 

For statistical homogeneous medium having translational symmetry, $\bm{\Sigma}(\mathbf{p},\mathbf{p}')=\bm{\Sigma}(\mathbf{p})\delta(\mathbf{p}-\mathbf{p}')$ and $\langle\mathbf{G}(\mathbf{p},\mathbf{p}')\rangle=\langle\mathbf{G}(\mathbf{p})\rangle\delta(\mathbf{p}-\mathbf{p}')$ \cite{sheng2006introduction}. In the momentum representation the free-space dyadic Green's function is $ \mathbf{G}_{0}(\mathbf{p})=-{1}/(k^{2}\mathbf{I}-p^{2}(\mathbf{I}-\mathbf{\hat{p}}\mathbf{\hat{p}}))$, the averaged amplitude Green's function is then 
\begin{equation}
\langle\mathbf{G}(\mathbf{p})\rangle=-\frac{1}{k^{2}\mathbf{I}-p^{2}(\mathbf{I}-\mathbf{\hat{p}}\mathbf{\hat{p}})-\bm{\Sigma}(\mathbf{p})} 
\end{equation}
where $\mathbf{\hat{p}}=\mathbf{p}/p$ is the unit vector in the momentum space. Through this equation, self-energy $\bm{\Sigma}(\mathbf{p}) $ provides a renormalization for the electromagnetic wave propagation in random media, and determines the effective (renormalized) permittivity as
\begin{equation}\label{epsilon_eff}
\bm{\varepsilon}_{eff}(\mathbf{p})=\mathbf{I}-\frac{\bm{\Sigma}(\mathbf{p})}{k^2}
\end{equation} 
For a statistically isotropic random medium, the obtained momentum-dependent effective permittivity tensor is decomposed into a transverse part and a longitudinal part as $
\bm{\varepsilon}(\mathbf{p})=\varepsilon^{\bot}(\mathbf{p})(\mathbf{I}-\mathbf{\hat{p}}\mathbf{\hat{p}})+\varepsilon^{\parallel}(\mathbf{p})\mathbf{\hat{p}}\mathbf{\hat{p}}$, where $\varepsilon^{\bot}(\mathbf{p})=1-\Sigma^{\bot}(\mathbf{p})/{k^{2}}$ and $\varepsilon^{\parallel}(\mathbf{p})=1-\Sigma^{\parallel}(\mathbf{p})/{k^{2}}$ determine the effective permittivities of transverse and longitudinal modes in momentum space. Therefore, by determining the poles of amplitude Green's function we can obtain the dispersion relation which corresponds collective excitation of the disordered medium. For transverse waves, the dispersion relation is $\mathbf{K}^{ 2}=\mathbf{\varepsilon}^{\bot}(\mathbf{K})k^{2}$. $\mathbf{K}$ is viewed as the effective propagation wave vector for the disordered medium. 

It is noted that Dyson equation and self-energy only provide a characterization for coherent electromagnetic field propagation in random media, i.e., the first moment of electromagnetic field, while a more relevant quantity to our concern is the radiation intensity in random media which directly determines the phase function of each scattering process in terms of energy transport. This is exactly governed by the Bethe-Salpeter equation \cite{lagendijk1996resonant,VanRossum1998,tsang2004scattering}, which describes the second moment of the electromagnetic field $\langle \mathbf{G}\mathbf{G}^*\rangle$ in random media. In operator notation, Bethe-Salpeter equation is written as
\begin{equation}
\langle\mathbf{G}\mathbf{G}^*\rangle=\langle\mathbf{G}\rangle\langle\mathbf{G}^*\rangle+\langle\mathbf{G}\rangle\langle\mathbf{G^*}\rangle\bm{\Gamma}\langle\mathbf{G}\mathbf{G}^*\rangle
\end{equation} 
where $\bm{\Gamma}$ is the irreducible vertex representing the renormalized scattering center for incoherent part of radiation intensity due to random fluctuation of the disordered media. It can be understood as the differential scattering cross section related as well as scattering phase function in radiative transfer. However, it should be noted that the irreducible vertex is a much more complex quantity describing all interference phenomena in multiple scattering.
\subsection{Quasicrystalline approximation}
\begin{figure}[htbp]
	\flushleft
	\subfloat{
		\includegraphics[width=\linewidth]{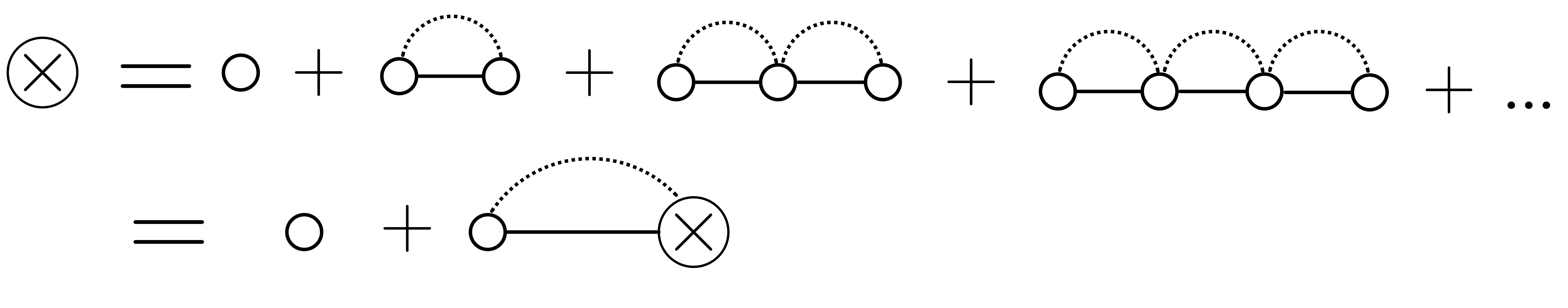}
	}
	\caption{Diagrammatic representation of $\bm{\Sigma}$ under quasicrystalline approximation (QCA). The hollow circle denotes single particle $t$-matrix, solid line denotes the free-space Green's function (propagator), dotted line denotes particle correlation function $h_2(r)=g_2(r)-1$, and the large circle with a $\times$ inside denotes $\bm{\Sigma}$.}\label{diagram_qca}
\end{figure}

The primary task at hand is to evaluate the self-energy to obtain an exact enough ensemble-averaged Green's function. As a first-order perturbative approximation, under independent scattering approximation (ISA) self-energy is simply the configurational sum of the $t$-matrices for all scatterers 
\begin{equation}
\bm{\Sigma}^{(1)}(\mathbf{p})=n_0t\mathbf{I} 
\end{equation}
where $t=-k^2\alpha$ is the \textit{t}-matrix of the dipole scatterer. In ISA, all particles scatter light independently as if no other particles exist. This approximation is only valid for very dilutely distributed scatterers. In denser random media, dependent scattering mechanism originating for wave interference arises. Since the scatterers are randomly distributed in the medium and have finite sizes comparable with wavelength, it is pivot to take the inter-particle correlation into account in the dependent scattering mechanism \cite{fradenPRL1990,rojasochoaPRL2004}. This is because the existence of one particle would create an exclusion volume into which other particles are not allowed to penetrate, if we regard the particles as hard spheres, leading to definite phase differences among scattered waves preserving over ensemble average. Here we only consider the correlation between a pair of particles, which is described by the pair distribution function $g_2(\mathbf{r}_1,\mathbf{r}_2)=g_2(|\mathbf{r}_1-\mathbf{r}_2|)$ of the hard-sphere type inter-particle correlation, where we have implicitly assumed the random medium is statistically homogeneous and isotropic. $g_2(r)$ describes the probability of finding a particle in the distance of $r$ from a fixed particle. High-order position correlations involving three or more particles simultaneously are treated as a hierarchy of pair distribution functions, e.g., $g_3(\mathbf{r}_1,\mathbf{r}_2,\mathbf{r}_3)=g_2(\mathbf{r}_1,\mathbf{r}_2)g_2(\mathbf{r}_2,\mathbf{r}_3)$, $g_4(\mathbf{r}_1,\mathbf{r}_2,\mathbf{r}_3,\mathbf{r}_4)=g_2(\mathbf{r}_1,\mathbf{r}_2)g_2(\mathbf{r}_2,\mathbf{r}_3)g_2(\mathbf{r}_3,\mathbf{r}_4)$ and so forth. This approximating method allows us to solve the propagation problem of coherent electromagnetic field (or mean field) in random media, and is called quasicrystalline approximation (QCA) \cite{laxPR1952,tsangJAP1982}. The diagrammatic representation of self-energy under QCA is presented in Fig.\ref{diagram_qca}, which only contains irreducible diagrams of multiple scattering, as we have mentioned in Section \ref{general_theory}.

In low-frequency limit for dipole particles, the self-energy under QCA is expressed in a self-consisted way in reciprocal space according to Fig.\ref{diagram_qca} as
\begin{equation}
\bm{\Sigma}\left(\mathbf{p} \right)=n_0t \mathbf{I}+n_0t\int\mathbf{G}_0\left( \mathbf{p} \right)H_2\left( \mathbf{p}\right)\Sigma\left(\mathbf{p} \right) 
\end{equation}

Let $\bm{\Sigma}\left(\mathbf{p} \right)=\Sigma\mathbf{I}$ be $\mathbf{p}$ independent, which is applicable for small particles meaning the the scattering properties are local with no spatial dispersion and we have
\begin{equation}\label{qca1}
\Sigma\mathbf{I}=n_0t\mathbf{I}-\frac{n_0t\Sigma}{3k^2}\mathbf{I}+n_0t\Sigma\int{d\mathbf{r}\mathrm{PV}\mathbf{G}_0\left( \mathbf{r} \right)h_2(r)}
\end{equation}
where we have split the singular part of Green's function and define the principal value of Green's function as $\mathrm{PV}\mathbf{G}_0\left( \mathbf{r} \right)$. The integral is also transformed from reciprocal domain to space domain. Then self-energy is solved as 
\begin{equation}
\Sigma=\frac{n_0t}{1+n_0t/(3k^2)+2n_0t\int{drr\exp(ikr)\left[g_2(r)-1\right]}/3}
\end{equation}
Therefore according to Eq.(\ref{epsilon_eff}) the effective propagation constant $K$ is given by 
\begin{equation}
K^2=k^2-\frac{1}{1/(n_0t)+1/(3k^2)+2\int{drr\exp(ikr)\left[g_2(r)-1\right]}/3}
\end{equation}
\begin{figure}[htbp]
	\flushleft
	\subfloat{
		\includegraphics[width=\linewidth]{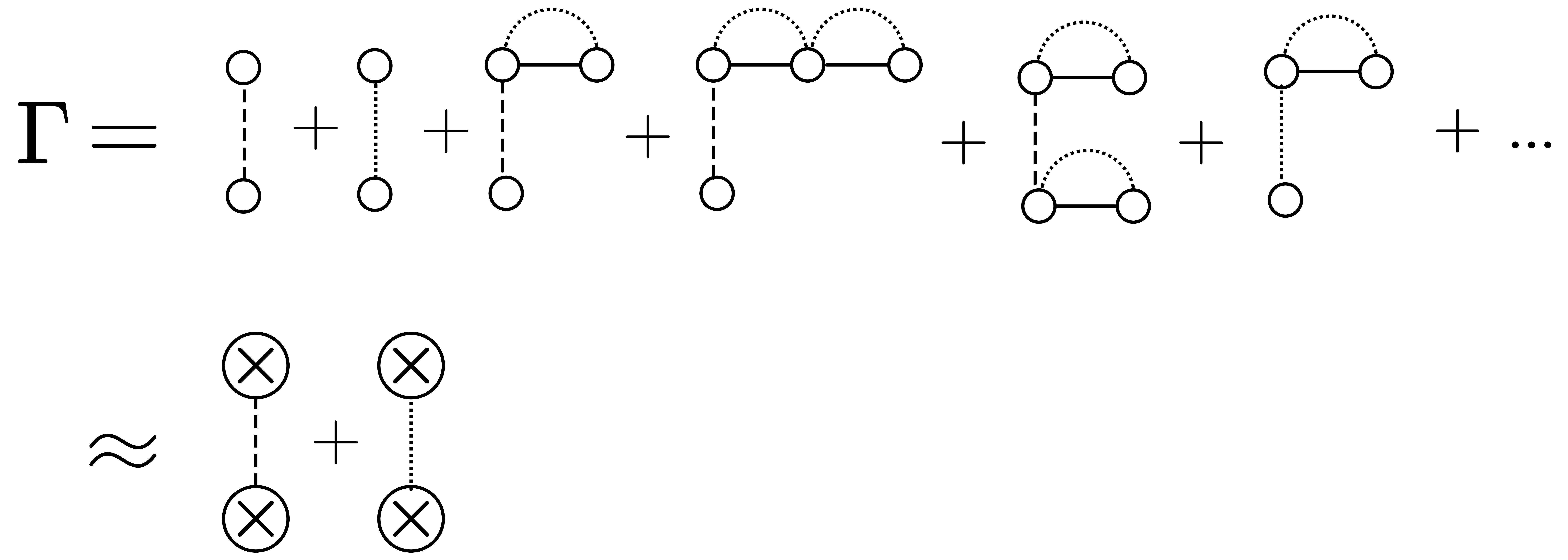}
	}
	\caption{Diagrammatic representation of irreducible intensity vertex $\bm{\Gamma}$ under QCA. The dashed lines denote the connected particles are the same one.}\label{gamma_diagram}
\end{figure}

After calculating the effective propagation constant for the coherent wave, we then proceed to derive the differential scattering cross section for incoherent waves. It is determined by the irreducible intensity vertex $\Gamma$. Its diagrammatic representation under QCA is also given in Fig.\ref{gamma_diagram}. Again here we only consider two-particle statistics. Note the derivation can be simplified using $\Sigma$ in the second line of Fig.\ref{gamma_diagram} is due to our local assumption for the scattering process, which reduces the correlation function into only two particle positions. This simplification is important and is to some extent equivalent to the so-called ``distorted Born approximation''\cite{maAO1988,tsangRS2000,tsang2004scattering}. The irreducible (intensity) vertex is solved as
\begin{equation}\label{gamma_qca}
\Gamma(\mathbf{p},\mathbf{p}')=n_0|c|^2+n_0^2|c|^2H_2(\mathbf{p}-\mathbf{p}')
\end{equation}
where $c=\Sigma/n_0$, and $H_2(\mathbf{q})$ is defined as the Fourier transform of pair correlation function $h_2(r)=g_2(r)-1$ as
\begin{equation}\label{Hq_eq}
H_2(\mathbf{q})=\int_{-\infty}^{\infty}d^3\mathbf{r}[g_2(r)-1]\exp{(-i\mathbf{q}\cdot\mathbf{r})}
\end{equation}

Afterwards, we take the on-shell approximation, which implies the photons transport with a fixed momentum value $p=K$ and those excitations with other momentum values are negligible, like a shell in the momentum (reciprocal) space. This approximation also corresponds to a mean Green's function with a sharp Dirac-type peak at $p=K$. It gives
\begin{equation}\label{gamma_qca2}
\Gamma(K\hat{\mathbf{p}},K\hat{\mathbf{p}}')=n_0|c|^2+n_0^2|c|^2H_2(K\hat{\mathbf{p}}-K\hat{\mathbf{p}}')
\end{equation}

Since we only consider transverse electromagnetic waves, the the transverse component of irreducible vertex plays the role of renormalized scattering center for transverse electromagnetic intensity, which is given by \cite{Cherroret2016}
\begin{equation}\label{gamma_qca3}
\Gamma^{\bot}(K\hat{\mathbf{p}},K\hat{\mathbf{p}}')=(\mathbf{I}-\hat{\mathbf{p}}\hat{\mathbf{p}})\Gamma(K\hat{\mathbf{p}},K\hat{\mathbf{p}}')(\mathbf{I}-\hat{\mathbf{p}}'\hat{\mathbf{p}}')
\end{equation}

For isotropic media, the scattering properties (scattering coefficient, phase function) do not rely on the incident direction but only the angle between the incident and scattering direction, i.e., the polar scattering angle $\theta_s=\arccos(\hat{\mathbf{p}}\cdot\hat{\mathbf{p}}')$ and azimuth angle $\varphi_s$, which depends on the definition of local frame of spherical coordinates with respect to the incident direction $\hat{\mathbf{p}}'$. Here the isotropy is manifested in the express of pair correlation function $H_2(K\hat{\mathbf{p}}-K\hat{\mathbf{p}}')$, which only depends on the difference between $\hat{\mathbf{p}}$ and $\hat{\mathbf{p}}'$. For unpolarized light transport in random media, azimuth symmetry is preserved. Therefore, the differential scattering cross section can be obtained by integrating over $\varphi_s$ with incident direction $\hat{\mathbf{p}}'$ fixed as \cite{barabanenkov1975multiple}
\begin{equation}\label{pf_qca1}
\begin{split}
\frac{d\sigma_s'}{d\theta_s}=\frac{1}{(4\pi)^2}\int_{0}^{2\pi}\Gamma^{\bot}(K\hat{\mathbf{p}},K\hat{\mathbf{p}}')d\varphi_s
\end{split}
\end{equation}
This leads to the following result of differential scattering cross section that is not dependent on the specific incident direction $\hat{\mathbf{p}}'$ but the relative polar scattering angle $\theta_s$ between $\hat{\mathbf{p}}$ and $\hat{\mathbf{p}}'$ as
\begin{equation}\label{pf_qca2}
\begin{split}
\frac{d\sigma_s'}{d\theta_s}=\frac{n_0|c|^2(1+\cos^2\theta_s)}{16\pi n_{eff}}\left[1+n_0H_2(2K\sin(\theta_s/2))\right]
\end{split}
\end{equation}
The effective refractive index $n_{eff}$ in the denominator indicates that  the differential scattering coefficient is with respect to the coherent transmitted energy flux, rather than the incident energy flux. It is given by the effective propagation constant as
\begin{equation}
n_{eff}=\frac{\mathrm{Re}K}{k}
\end{equation}
Therefore the scattering coefficient which is defined as scattering cross section per unit volume is given by\footnote{In the long-wavelength limit, the structure factor can be replaced by its $\mathbf{q}=0$ value, leading to
	$\mu_s'=n_0|c|^2S(\mathbf{q}=0)/(6\pi)$. However, this is not valid for the present case with $k_0a\sim 0.6.$}
\begin{equation}\label{kappas_qca}
\begin{split}
\mu_s'&=\int_0^\pi\frac{d\sigma_s'}{d\theta_s}\sin{\theta_s}d\theta=\int_0^\pi d\theta\sin{\theta_s}\frac{n_0|c|^2(1+\cos^2\theta_s)}{16\pi n_{eff}}\\&\times\left[1+n_0H_2(2K\sin(\theta_s/2))\right]
\end{split}
\end{equation}

Since the spheres are all hard, impenetrable spheres without any other interparticle potentials, we can use the Percus-Yevick hard-sphere model for the pair correlation function \cite{Baxter1968,tsang2004scattering2}. In this model, $H_2(\mathbf{q})$ is calculated as 
\begin{equation}
H_2(\mathbf{q})=\frac{(2\pi)^3C(\mathbf{q})}{1-n_0(2\pi)^3C(\mathbf{q})}
\end{equation}
where
\begin{equation}
\begin{split}
C(\mathbf{q})&=C(q)=24f_v[\frac{\alpha+\beta+\delta}{u^2}\cos u-\frac{\alpha+2\beta+4\delta}{u^3}\sin u\\&-2\frac{\beta+6\delta}{u^4}\cos u+\frac{2\beta}{u^4}+\frac{24\delta}{u^6}(\cos u-1)]
\end{split}
\end{equation}
in which $u=4qa$, $\alpha=(1+2f_v)^2/(1-f_v)^4$, $\beta=-6f_v(1+f_v/2)^2/(1-f_v)^4$, $\delta=f_v(1+2f_v)^2/[2(1-f_v)^2]$ and $f_v$ is the volume fraction of the identical spherical particles. 

However, the self-energy and irreducible vertex under QCA don't exactly fulfill the so-called Ward-Takahashi identity for energy conservation \cite{tsang2004scattering}, hence we use another method to determine the absorption coefficient. Since in the present case the self-energy is local and independent of wavevector, the quantity $c$ can be understood as an effective scattering center or a renormalized scatterer. This quantity gives a modification for the exciting field impinging on the scatterers as $c/t$. In this way, the absorption coefficient is calculated from the single particle absorption cross section as
\begin{equation}
\mu_a'=\frac{n_0}{n_{eff}}\Big|\frac{c}{t}\Big|^2\left(-\frac{\mathrm{Im}t}{k}-\frac{|t|^2}{6\pi}\right)
\end{equation}
The quantity in the parentheses is the single scatterer absorption cross section. Again the effective refractive index $n_{eff}$ in the denominator plays the same role as in Eq.(\ref{pf_qca1}). 

Therefore, we have established a dependent scattering model for radiative properties of randomly distributed dipole scatterers based on QCA and multiple scattering theory. This is the main theoretical contribution of the present paper. This model is not equivalent to the conventional QCA in the low-frequency limit as used by many authors \cite{ma1990enhanced,wen1990dense,prasherJAP2007,weiAO2012}. In this model, thanks to the diagrammatic technique in multiple scattering theory, we can formally and rigorously derive the irreducible intensity vertex under QCA as well as the dependent absorption coefficient, and take the anisotropic scattering phase function into account. It is thus still applicable for larger dipolar particles with size parameter $x=ka$ approaching 1, in which particle correlation plays a more important role in affecting radiative properties than the case of very small particles. 
\section{Results and Discussions}\label{results}
\subsection{Dependent scattering effect on radiative properties}
\begin{figure}[htbp]
	%\captionsetup{labelformat=simple}
	\includegraphics[width=0.8\linewidth]{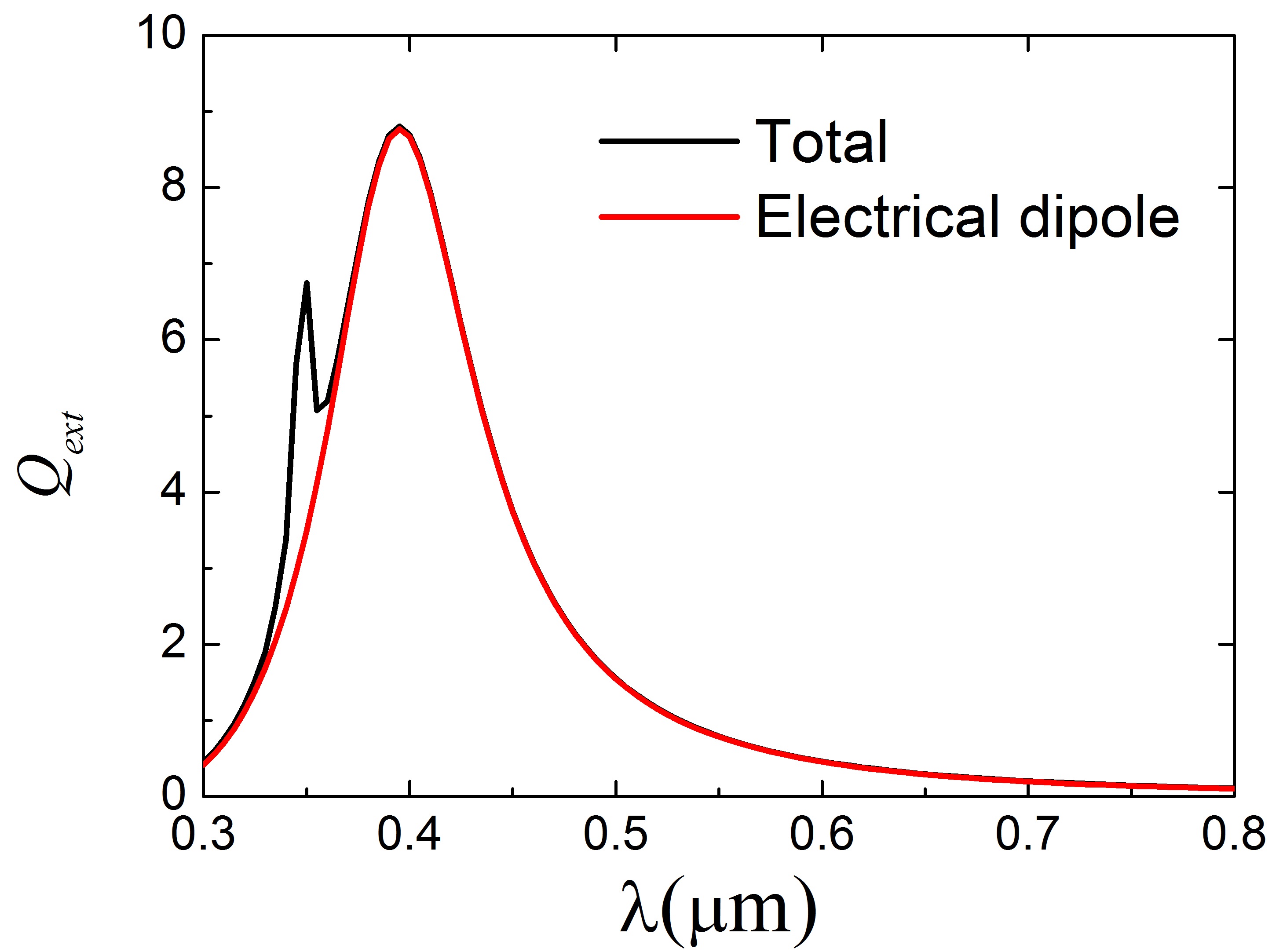}
	\caption{Extinction efficiency $Q_{ext}$ of a single silver spherical particle with  radius $a=50nm$ at different wavelengths, in which the full Mie theory result (black solid line) is compared with the electric dipole approximation (red solid line).}\label{singleqext}
\end{figure}
Here we investigate a random medium consisting of identical silver spheres and their radius is set to be $a=50$nm, a typical size of silver nanoparticles. The metallic particle made of Ag, whose permittivity is described by the Drude formula
\begin{equation}
\varepsilon(\omega)=\varepsilon_r-0.73\frac{\omega_p^2}{\omega(\omega+i\gamma)}
\end{equation}
with $\varepsilon_r=5.45$, $\omega_p=1.72\times10^{16}\text{rad/s}$ and decay rate $\gamma=8.35\times10^{13}\text{rad/s}$ \cite{shore2012complex}.

The total extinction efficiency defined as $Q_{ext}=C_{ext}/(\pi a^2)$ is calculated through Mie theory to take high orders of multipoles into account, shown in Fig.\ref{singleqext}. The extinction efficiency under electric dipole approximation is also calculated to be $Q_{ext}=k\mathrm{Im}\alpha/(\pi a^2)$, where $\alpha$ is electric dipole polarizability obtained from Eq.(\ref{alphamie}). It can be observed that for wavelength $\lambda>0.35\mu m$, the electrical dipole approximation is able to describe silver nanoparticle's single scattering properties. Therefore, except for very dense systems where near-field electromagnetic interaction can induce the coupling of high order multipole modes between different particles, the present medium can be well described by a cluster of electric dipoles. In this condition, the numerical CDM can be seen as an exact treatment for electromagnetic wave propagation in such medium, and our theoretical model can be applied.

\begin{figure}[htbp]
	%\captionsetup{labelformat=simple}
	\subfloat{
		\label{musca}
		\includegraphics[width=0.8\linewidth]{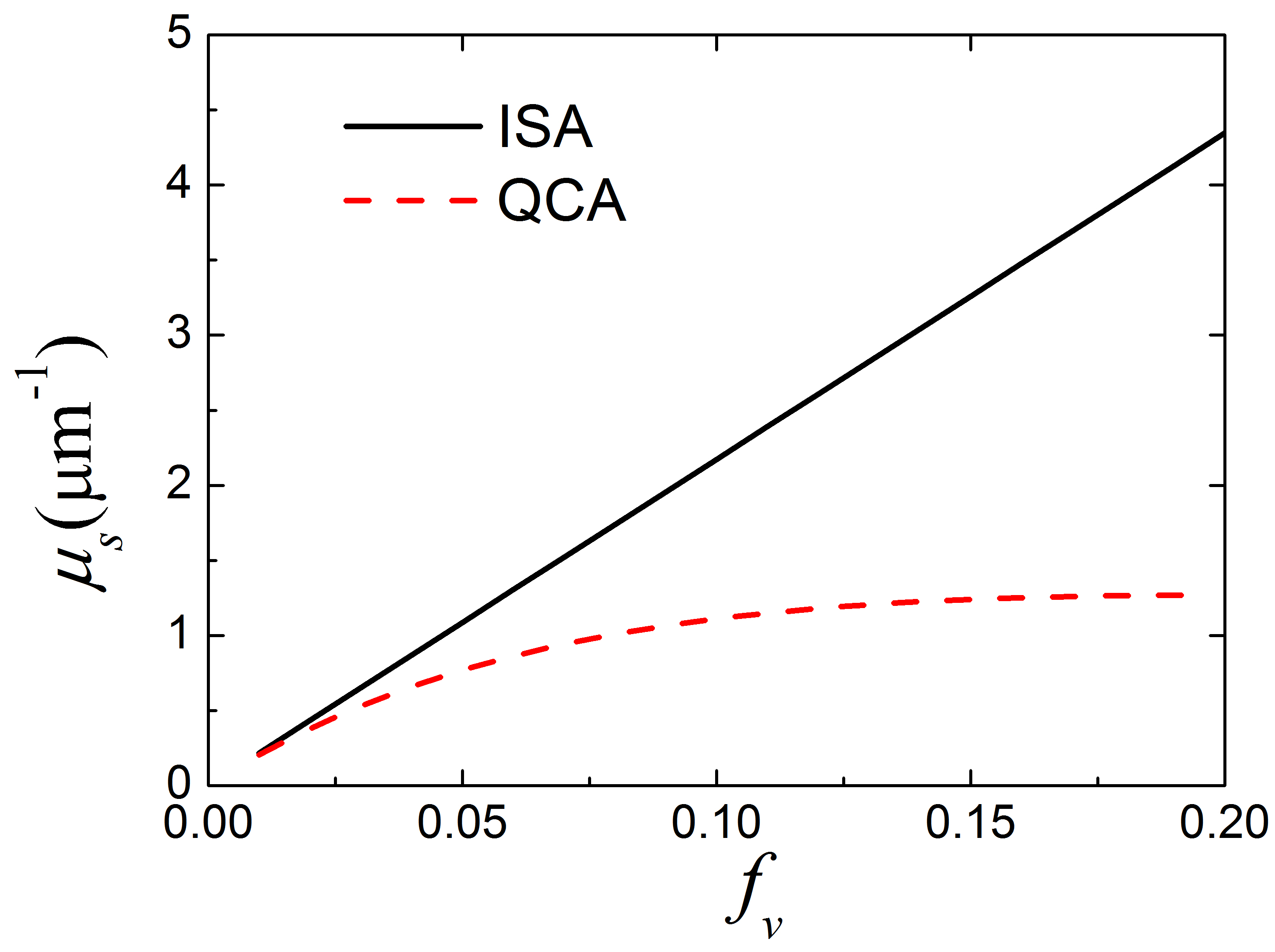}
	}
	\hspace{0.01in}
	\subfloat{
		\label{muabs}
		\includegraphics[width=0.8\linewidth]{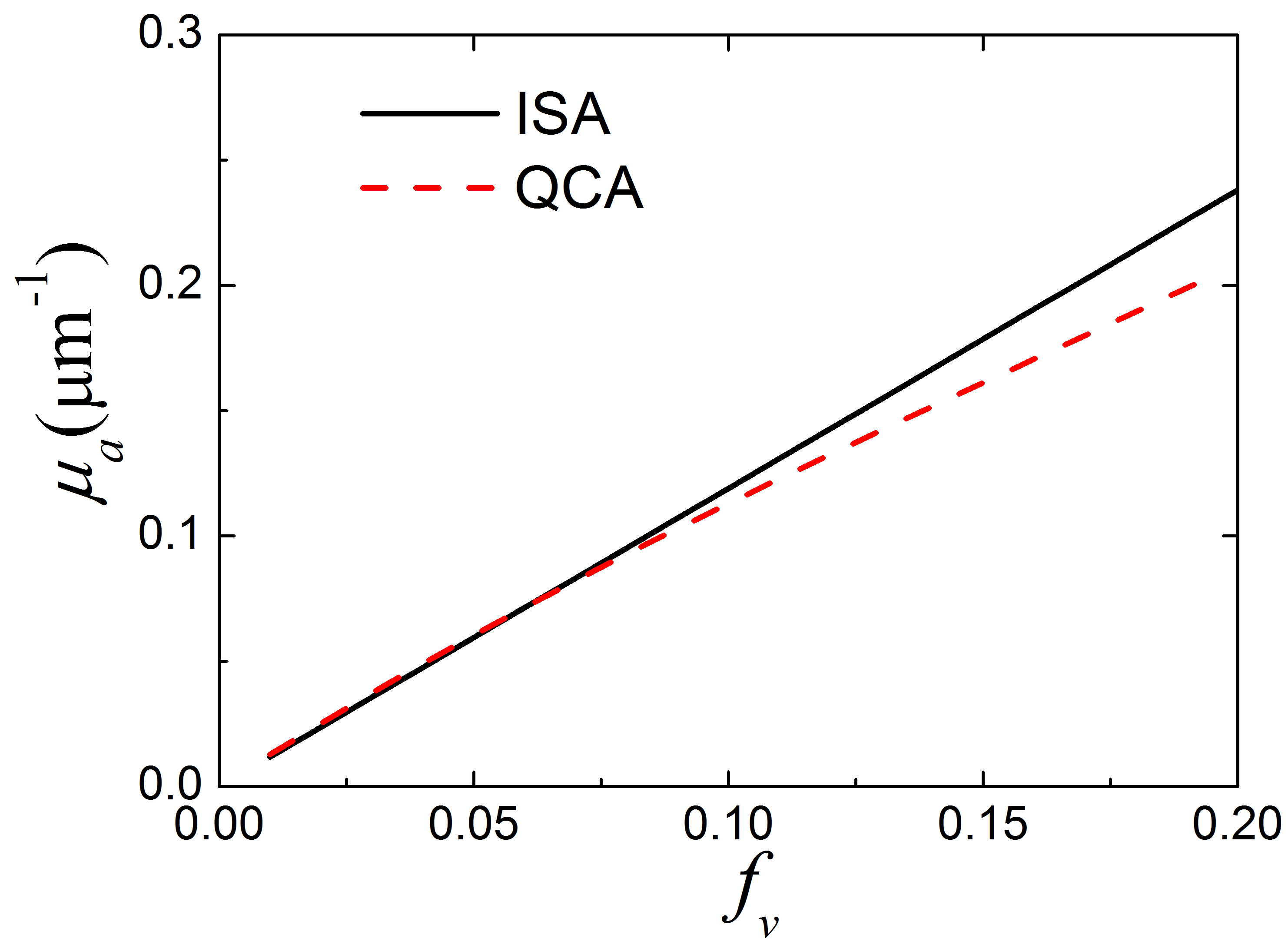}
	}
	\caption{ Radiative properties (a) scattering coeffcient $\mu_s$ (b) absorption coefficient $\mu_a$ calculated from independent scattering approximation (ISA) and quasicrystalline approximation (QCA) for different volume fractions $f_v$ of nanoparticles.}\label{model_properties}
\end{figure}

Here we focus on the operating wavelength of $\lambda=0.5\mu m$, where the size parameter of nanoparticle $x=2\pi a/\lambda=0.628$. A single silver particle has scattering and absorption efficiencies of $Q_{sca}=1.4596$ and $Q_{abs}=0.0929$ respectively, showing a highly scattering and weakly absorptive feature. The scattering and absorption coefficients of the random media with different nanoparticle volume fractions calculated using ISA and QCA are then presented in Fig.\ref{model_properties}. Even though $x$ is not much smaller than 1 as assumed in our model, we expect our theoretical model assuming local radiative properties to still be applicable. In Fig.\ref{model_properties}, it is obvious that ISA exhibits a linear increase of scattering and absorption coefficient with volume fraction, while QCA predicts a much smaller scattering coefficient when the volume fraction increases. The two models only overlap at very low volume fraction ($f_v\sim0.01$). We only investigate volume fraction up to $0.2$ since for higher densities, two-particle statistics only is not able to correctly capture multi-particle correlations in random media, making QCA invalid \cite{westJOSAA1994}. When $f_v=0.2$, scattering coefficient under ISA is almost four times larger than that of QCA, demonstrating that dependent scattering, in the present case, hugely reduces scattering coefficient. On the other hand, the difference of absorption coefficient between the two models is less significant, where dependent scattering mechanism slightly reduces absorption coefficient.

\begin{figure}[htbp]
	%\captionsetup{labelformat=simple}
	\subfloat{
		\label{asymfac}
		\includegraphics[width=0.8\linewidth]{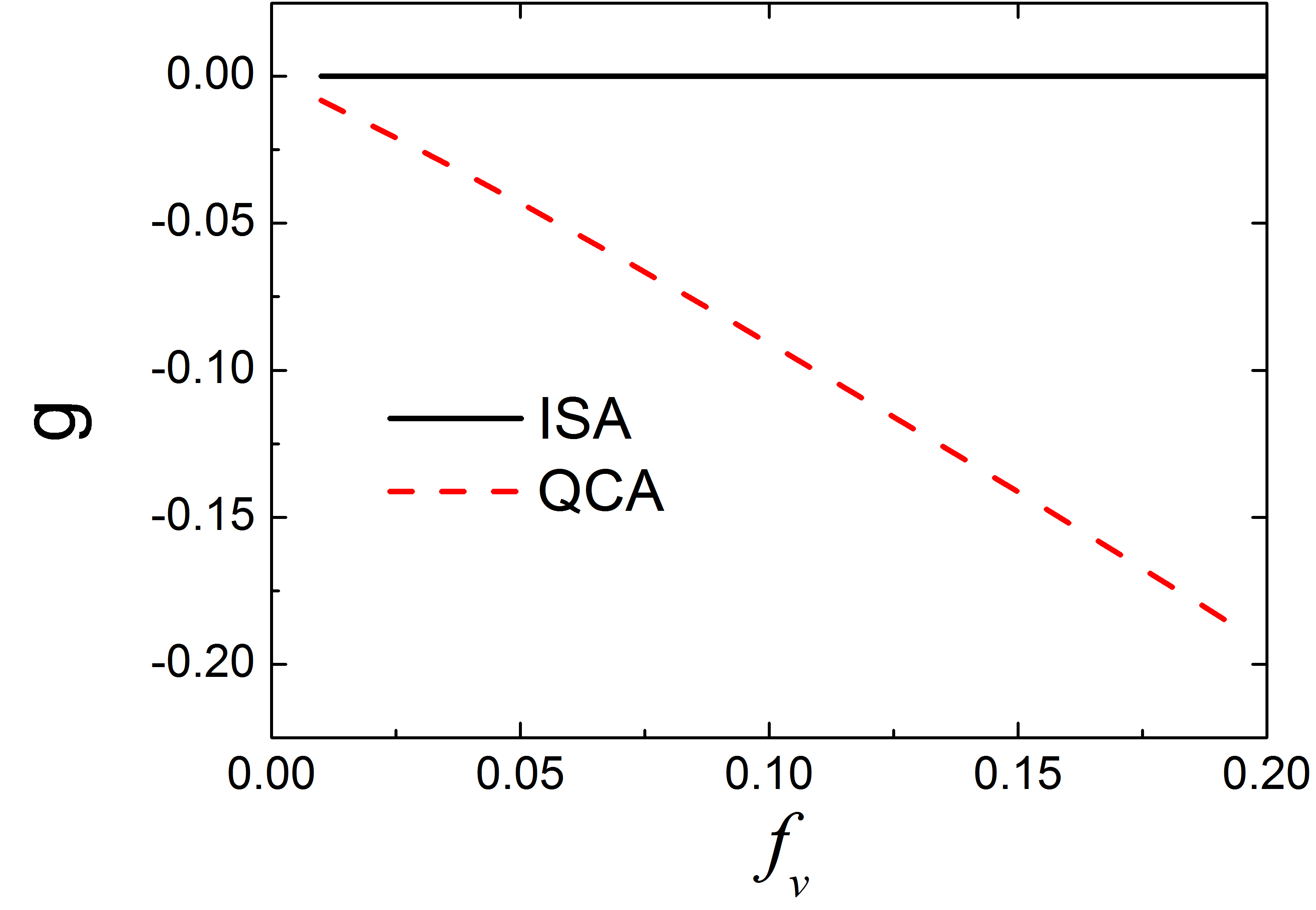}
	}
	\hspace{0.01in}
	\subfloat{
		\label{phasefunc}
		\includegraphics[width=0.8\linewidth]{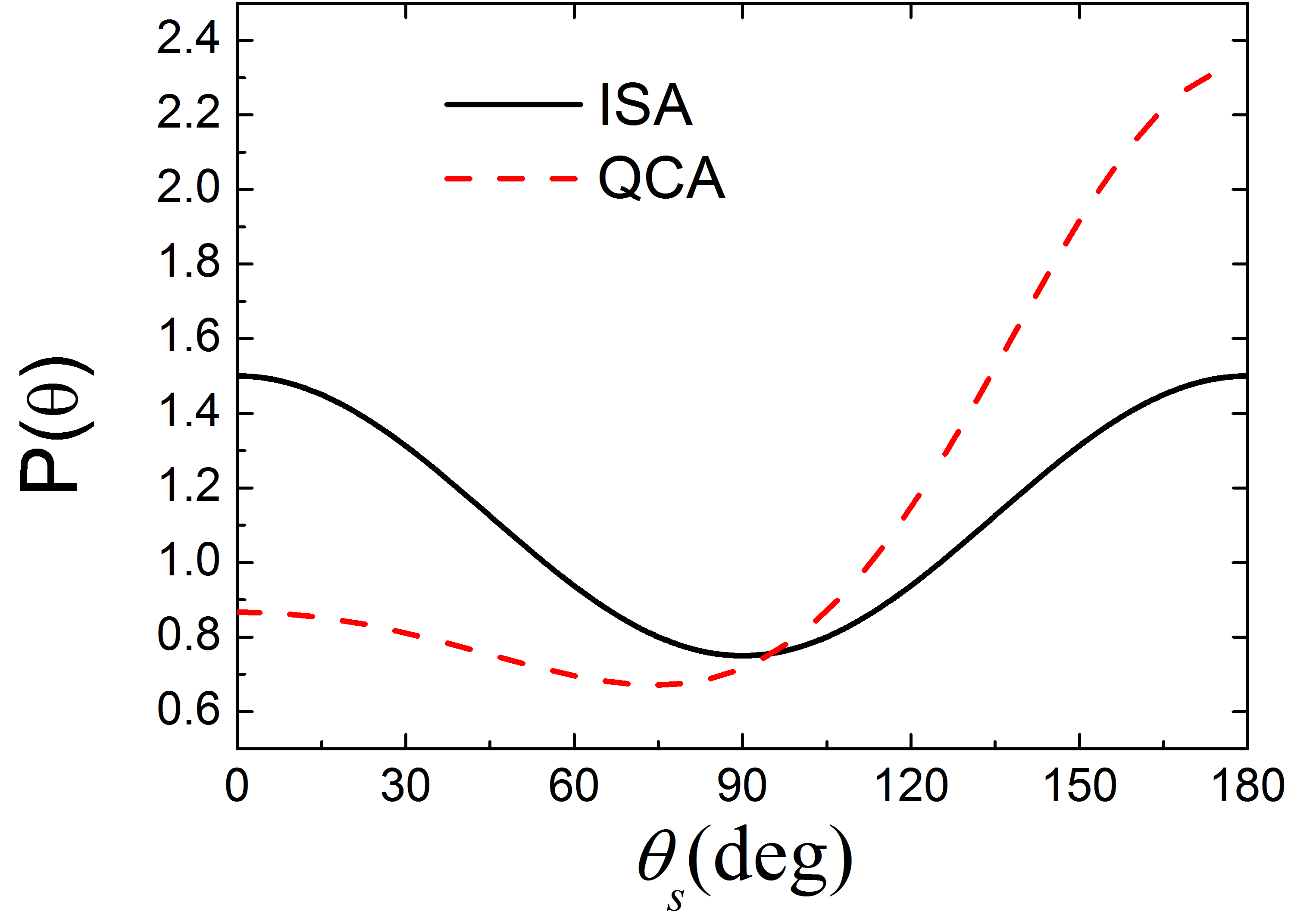}
	}
	\caption{(a) Asymmetry factor $g$ calculated by QCA for different volume fractions $f_v$ of nanoparticles, where for ISA $g$ is constantly zero. (b) Phase function under ISA and QCA for the $f_v=0.2$ case.}\label{anisotropy}
\end{figure}
Another feature of dependent scattering is to alter the scattering phase function. In Fig.\ref{asymfac} we show the scattering asymmetry factor $g$, defined as the mean cosine of scattering phase function, predicted by ISA and QCA varying with volume fraction. In ISA, the scattering phase function is always Rayleigh type and thus $g$ is kept to be zero. In the meanwhile, QCA gives rise to a negative $g$, which further decreases when volume fraction of particles grows. This suggests that dependent scattering enhances backscattering of radiation in the present medium. This feature arises from the structural correlation between particles, which enters into the phase function in the form of structure factor $S(\theta_s)1+n_0H_2(2K\sin(\theta_s/2))$ in Eq.(\ref{pf_qca2}). It is usually called by many authors as ``short-range order induced Bragg scattering'' in backscattering direction \cite{rojasochoaPRL2004}, which recently finds its application in structural coloration \cite{magkiriadou2012disordered}. In Fig.\ref{phasefunc}, we further plot the phase functions under ISA and QCA for $f_v=0.2$, where the backscattering enhancement is appreciable.
\subsection{Dependent scattering effect on total absorptance}
\begin{figure}[htbp]
	%\captionsetup{labelformat=simple}
	\includegraphics[width=0.8\linewidth]{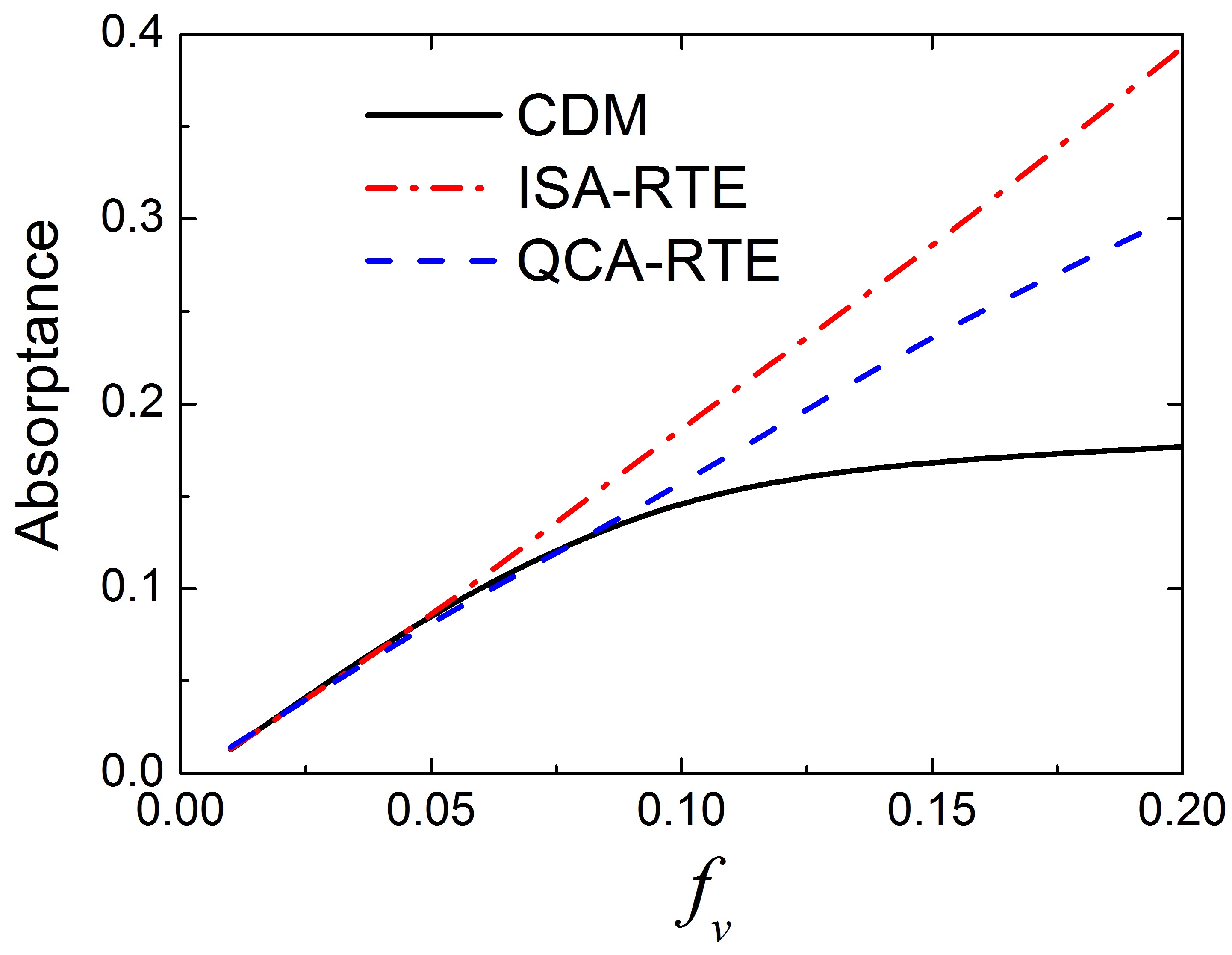}
	\caption{Variation of absorptance with nanoparticle volume fraction $f_v$ under a fixed slab thickness $L_z=1 \mathrm{\mu m}$.}\label{absorptance1}
\end{figure}
To verify the effectiveness of QCA model for radiative properties, we insert the calculated scattering and absorption coefficients and phase function into RTE to obtain the total absorptance of the random medium slab. We denote the combined modeling as ISA-RTE and QCA-RTE. In the following calculation we have fixed $L_x=L_y$ to be $5\mathrm{\mu m}$, which are sufficiently large than the operating wavelength ($\lambda=0.5\mu m$) in order to reduce side diffraction effects. The thickness of the slab, $L_z$, is varied from $L_z=0.1\mathrm{\mu m}$ to $L_z=2\mathrm{\mu m}$ to cover a sufficiently broad optical thicknesses from a monolayer of particles to optically thick slabs. The incident light propagates along the direction $z$ as shown in Fig.\ref{system_config}. 

The results are then compared with the full wave simulations of CDM. We first fix the thickness of the slab to be $L_z=1\mu m$ and vary the volume fraction of nanoparticles, and the comparison is shown in Fig.\ref{absorptance1}. It can be observed that when $f_v<0.05$, the three methods show almost the same result. When $f_v$ continues to grows, ISA starts to deviate and QCA can still predict correct absorptance. However, when $f_v>0.1$, both methods are not capable to exactly reproduce the CDM results. Nevertheless, in high density random media, QCA can give much more reasonable predictions on absorptance than ISA.

\begin{figure}[htbp]
	%\captionsetup{labelformat=simple}
	\includegraphics[width=0.8\linewidth]{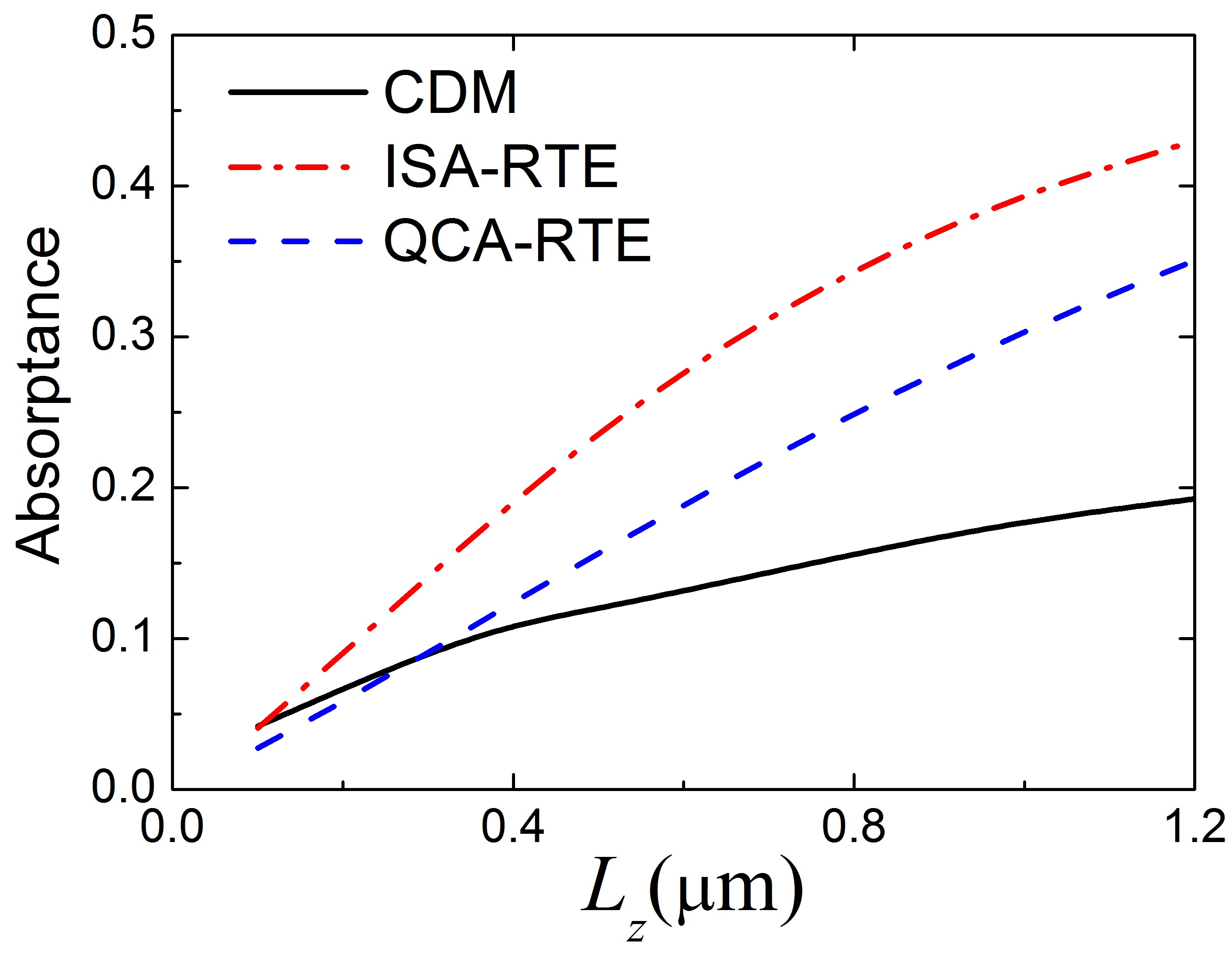}
	\caption{Variation of absorptance with slab thickness $L_z$ under a fixed nanoparticle volume fraction $f_v=0.2$.}\label{absorptance2}
\end{figure}

In Fig.\ref{absorptance2}, we show the thickness dependence of absorptance for the three methods, where the thickness $L_z$ varies from $0.1\mathrm{\mu m}$ to $1.2\mathrm{\mu m}$, and the volume fraction of particles is chosen to be $f_v=0.2$. It is found that when $L_z=0.1\mathrm{\mu m}$, ISA-RTE converges to CDM while QCA-RTE deviates. This is because in the particle monolayer, multiple scattering of electromagnetic waves is weak and the absorptance is dominated by single scattering regime. On the other hand, our QCA model is established for a sufficiently large random medium where Dyson and Bethe-Salpeter equations describing multi-particle scattering processes apply, and it is actually invalid and leads to errors for such monolayer. When $L_z$ increases, it is seen that QCA results in more closer absorptance to CDM than ISA in the entire thickness range, especially for $L_z<0.4\mathrm{\mu m}$. However, the deviations of both models are still very large for thicker slabs and further grow with thickness. This indicates that for very dense, highly scattering random media with appreciable optical thickness (under QCA, for $L_z=0.4\mathrm{\mu m}$, the optical thickness $\tau=L_z(\mu_a+\mu_s)=0.591$)  where multiple scattering mechanism prevails, QCA model is still not sufficient to predict total absorptance, although it already indicates much better performance than ISA. 

Particularly, in such thick and dense random media, high particle density brings several mechanisms that QCA can't capture or even neglect. The first mechanism is high-order correlation among three or more particles, which QCA treats approximately using a cascading multiplication of two particle statistics. The second is the recurrent scattering mechanism, which denotes the multiple scattering trajectories visit the same particle more than once and is ignored in QCA. Moreover, high optical thickness may also lead to very long multiple scattering paths that is responsible for the position dependent radiative properties, for example, position-dependent diffusion constant as investigated by many authors \cite{tiggelenPRL2000,yamilovPRL2014}. Effects of these mechanisms on light absorption in random media need further exploration and are out the scope of the present paper.

\section{Conclusions}
In this paper we theoretically and numerically demonstrate the effect of dependent scattering on absorption in disordered media consisting of highly scattering scatterers. Based on the path-integral diagrammatic technique in multiple scattering theory, we develop a theoretical model using quasi-crystalline approximation (QCA) to derive dependent-scattering corrected radiative properties. We investigate a disordered medium composed of randomly distributed silver nanoparticles with radius $a=50\text{nm}$ operating at $\lambda=0.5\mathrm{\mu m}$ with large single scattering cross sections and very small absorption cross sections. Compared to independent scattering approximation (ISA), we theoretically find that dependent scattering mechanism strongly reduces the overall scattering coefficient, and slight lowers the absorption coefficient. This reduction grows as the volume fraction of nanoparticles increases. Moreover, we demonstrate that dependent scattering leads to a backscattering enhancement in scattering phase function due to the local short-range correlations between particles. 

We further reveal that all these effects of dependent scattering on radiative properties will be reflected in the total absorptance of the random media. By performing comparison among the full-wave coupled-dipole method (CDM) (regarded as an exact benchmark in the present case), QCA-RTE and ISA-RTE, we find that dependent scattering effect shows a strong impact on total absorptance. When $f_v<0.05$, the three methods show almost the same result. When $f_v$ continues to grow, ISA-RTE starts to deviate and QCA-RTE can still predict correct absorptance. When $f_v>0.1$, both methods are not capable to exactly reproduce the CDM results. However, QCA-RTE still shows much better performance in predicting absorptance than ISA in high-density random media. We thus conclude that QCA-RTE is a feasible method to correctly treat absorption in random media in the dependent scattering regime, if the particle density is not too high.
\section*{Acknowledgments}
We thank the financial support of the National Natural Science Foundation of China (Grant Nos.51636004 and 51476097); Shanghai Key Fundamental Research Grant (Grant No.16JC1403200); The Foundation for Innovative Research Groups of the National Natural Science Foundation of China (Grant No.51521004).

\bibliography{dep_scat_abs}

\begin{thebibliography}{68}%
\makeatletter
\providecommand \@ifxundefined [1]{%
 \@ifx{#1\undefined}
}%
\providecommand \@ifnum [1]{%
 \ifnum #1\expandafter \@firstoftwo
 \else \expandafter \@secondoftwo
 \fi
}%
\providecommand \@ifx [1]{%
 \ifx #1\expandafter \@firstoftwo
 \else \expandafter \@secondoftwo
 \fi
}%
\providecommand \natexlab [1]{#1}%
\providecommand \enquote  [1]{``#1''}%
\providecommand \bibnamefont  [1]{#1}%
\providecommand \bibfnamefont [1]{#1}%
\providecommand \citenamefont [1]{#1}%
\providecommand \href@noop [0]{\@secondoftwo}%
\providecommand \href [0]{\begingroup \@sanitize@url \@href}%
\providecommand \@href[1]{\@@startlink{#1}\@@href}%
\providecommand \@@href[1]{\endgroup#1\@@endlink}%
\providecommand \@sanitize@url [0]{\catcode `\\12\catcode `\$12\catcode
  `\&12\catcode `\#12\catcode `\^12\catcode `\_12\catcode `\%12\relax}%
\providecommand \@@startlink[1]{}%
\providecommand \@@endlink[0]{}%
\providecommand \url  [0]{\begingroup\@sanitize@url \@url }%
\providecommand \@url [1]{\endgroup\@href {#1}{\urlprefix }}%
\providecommand \urlprefix  [0]{URL }%
\providecommand \Eprint [0]{\href }%
\providecommand \doibase [0]{http://dx.doi.org/}%
\providecommand \selectlanguage [0]{\@gobble}%
\providecommand \bibinfo  [0]{\@secondoftwo}%
\providecommand \bibfield  [0]{\@secondoftwo}%
\providecommand \translation [1]{[#1]}%
\providecommand \BibitemOpen [0]{}%
\providecommand \bibitemStop [0]{}%
\providecommand \bibitemNoStop [0]{.\EOS\space}%
\providecommand \EOS [0]{\spacefactor3000\relax}%
\providecommand \BibitemShut  [1]{\csname bibitem#1\endcsname}%
\let\auto@bib@innerbib\@empty
%</preamble>
\bibitem [{\citenamefont {Tien}\ and\ \citenamefont
  {Drolen}(1987)}]{tien1987thermal}%
  \BibitemOpen
  \bibfield  {author} {\bibinfo {author} {\bibfnamefont {C.-L.}\ \bibnamefont
  {Tien}}\ and\ \bibinfo {author} {\bibfnamefont {B.}~\bibnamefont {Drolen}},\
  }\href@noop {} {\bibfield  {journal} {\bibinfo  {journal} {Annual Review of
  Heat Transfer}\ }\textbf {\bibinfo {volume} {1}} (\bibinfo {year}
  {1987})}\BibitemShut {NoStop}%
\bibitem [{\citenamefont {Modest}(2013)}]{modest2013radiative}%
  \BibitemOpen
  \bibfield  {author} {\bibinfo {author} {\bibfnamefont {M.~F.}\ \bibnamefont
  {Modest}},\ }\href@noop {} {\emph {\bibinfo {title} {Radiative heat
  transfer}}}\ (\bibinfo  {publisher} {Academic press},\ \bibinfo {year}
  {2013})\BibitemShut {NoStop}%
\bibitem [{\citenamefont {Wiersma}(2013)}]{wiersma2013disordered}%
  \BibitemOpen
  \bibfield  {author} {\bibinfo {author} {\bibfnamefont {D.~S.}\ \bibnamefont
  {Wiersma}},\ }\href@noop {} {\bibfield  {journal} {\bibinfo  {journal} {Nat.
  Photon.}\ }\textbf {\bibinfo {volume} {7}},\ \bibinfo {pages} {188} (\bibinfo
  {year} {2013})}\BibitemShut {NoStop}%
\bibitem [{\citenamefont {Rotter}\ and\ \citenamefont
  {Gigan}(2017)}]{Rotter2017}%
  \BibitemOpen
  \bibfield  {author} {\bibinfo {author} {\bibfnamefont {S.}~\bibnamefont
  {Rotter}}\ and\ \bibinfo {author} {\bibfnamefont {S.}~\bibnamefont {Gigan}},\
  }\href {\doibase 10.1103/RevModPhys.89.015005} {\bibfield  {journal}
  {\bibinfo  {journal} {Rev. Mod. Phys.}\ }\textbf {\bibinfo {volume} {89}},\
  \bibinfo {pages} {015005} (\bibinfo {year} {2017})}\BibitemShut {NoStop}%
\bibitem [{\citenamefont {Tsang}\ and\ \citenamefont
  {Chang}(2000)}]{tsangRS2000}%
  \BibitemOpen
  \bibfield  {author} {\bibinfo {author} {\bibfnamefont {L.}~\bibnamefont
  {Tsang}}\ and\ \bibinfo {author} {\bibfnamefont {T.~C.}\ \bibnamefont
  {Chang}},\ }\href {\doibase 10.1029/1999RS002270} {\bibfield  {journal}
  {\bibinfo  {journal} {Radio Science}\ }\textbf {\bibinfo {volume} {35}},\
  \bibinfo {pages} {731} (\bibinfo {year} {2000})}\BibitemShut {NoStop}%
\bibitem [{\citenamefont {Mishchenko}\ \emph {et~al.}(2006)\citenamefont
  {Mishchenko}, \citenamefont {Travis},\ and\ \citenamefont
  {Lacis}}]{mishchenko2006multiple}%
  \BibitemOpen
  \bibfield  {author} {\bibinfo {author} {\bibfnamefont {M.~I.}\ \bibnamefont
  {Mishchenko}}, \bibinfo {author} {\bibfnamefont {L.~D.}\ \bibnamefont
  {Travis}}, \ and\ \bibinfo {author} {\bibfnamefont {A.~A.}\ \bibnamefont
  {Lacis}},\ }\href@noop {} {\emph {\bibinfo {title} {Multiple scattering of
  light by particles: radiative transfer and coherent backscattering}}}\
  (\bibinfo  {publisher} {Cambridge University Press},\ \bibinfo {year}
  {2006})\BibitemShut {NoStop}%
\bibitem [{\citenamefont {Xiao}\ \emph {et~al.}(2017)\citenamefont {Xiao},
  \citenamefont {Hu}, \citenamefont {Wang}, \citenamefont {Li}, \citenamefont
  {Tormo}, \citenamefont {Le~Thomas}, \citenamefont {Wang}, \citenamefont
  {Gianneschi}, \citenamefont {Shawkey},\ and\ \citenamefont
  {Dhinojwala}}]{xiaoSciAdv2017}%
  \BibitemOpen
  \bibfield  {author} {\bibinfo {author} {\bibfnamefont {M.}~\bibnamefont
  {Xiao}}, \bibinfo {author} {\bibfnamefont {Z.}~\bibnamefont {Hu}}, \bibinfo
  {author} {\bibfnamefont {Z.}~\bibnamefont {Wang}}, \bibinfo {author}
  {\bibfnamefont {Y.}~\bibnamefont {Li}}, \bibinfo {author} {\bibfnamefont
  {A.~D.}\ \bibnamefont {Tormo}}, \bibinfo {author} {\bibfnamefont
  {N.}~\bibnamefont {Le~Thomas}}, \bibinfo {author} {\bibfnamefont
  {B.}~\bibnamefont {Wang}}, \bibinfo {author} {\bibfnamefont {N.~C.}\
  \bibnamefont {Gianneschi}}, \bibinfo {author} {\bibfnamefont {M.~D.}\
  \bibnamefont {Shawkey}}, \ and\ \bibinfo {author} {\bibfnamefont
  {A.}~\bibnamefont {Dhinojwala}},\ }\href {\doibase 10.1126/sciadv.1701151}
  {\bibfield  {journal} {\bibinfo  {journal} {Science Advances}\ }\textbf
  {\bibinfo {volume} {3}} (\bibinfo {year} {2017}),\ 10.1126/sciadv.1701151},\
  \Eprint
  {http://arxiv.org/abs/http://advances.sciencemag.org/content/3/9/e1701151.full.pdf}
  {http://advances.sciencemag.org/content/3/9/e1701151.full.pdf} \BibitemShut
  {NoStop}%
\bibitem [{\citenamefont {Horstmeyer}\ \emph {et~al.}(2015)\citenamefont
  {Horstmeyer}, \citenamefont {Ruan},\ and\ \citenamefont
  {Yang}}]{horstmeyer2015guidestar}%
  \BibitemOpen
  \bibfield  {author} {\bibinfo {author} {\bibfnamefont {R.}~\bibnamefont
  {Horstmeyer}}, \bibinfo {author} {\bibfnamefont {H.}~\bibnamefont {Ruan}}, \
  and\ \bibinfo {author} {\bibfnamefont {C.}~\bibnamefont {Yang}},\ }\href@noop
  {} {\bibfield  {journal} {\bibinfo  {journal} {Nature photonics}\ }\textbf
  {\bibinfo {volume} {9}},\ \bibinfo {pages} {563} (\bibinfo {year}
  {2015})}\BibitemShut {NoStop}%
\bibitem [{\citenamefont {van Rossum}\ and\ \citenamefont
  {Nieuwenhuizen}(1998)}]{VanRossum1998}%
  \BibitemOpen
  \bibfield  {author} {\bibinfo {author} {\bibfnamefont {M.~C.~W.}\
  \bibnamefont {van Rossum}}\ and\ \bibinfo {author} {\bibfnamefont {T.~M.}\
  \bibnamefont {Nieuwenhuizen}},\ }\href {\doibase 10.1103/RevModPhys.71.313}
  {\bibfield  {journal} {\bibinfo  {journal} {Rev. Mod. Phys.}\ }\textbf
  {\bibinfo {volume} {71}},\ \bibinfo {pages} {86} (\bibinfo {year}
  {1998})}\BibitemShut {NoStop}%
\bibitem [{\citenamefont {Tsang}\ and\ \citenamefont
  {Kong}(2004)}]{tsang2004scattering}%
  \BibitemOpen
  \bibfield  {author} {\bibinfo {author} {\bibfnamefont {L.}~\bibnamefont
  {Tsang}}\ and\ \bibinfo {author} {\bibfnamefont {J.~A.}\ \bibnamefont
  {Kong}},\ }\href@noop {} {\emph {\bibinfo {title} {Scattering of
  Electromagnetic Waves: Advanced Topics}}}\ (\bibinfo  {publisher} {John Wiley
  \& Sons},\ \bibinfo {year} {2004})\BibitemShut {NoStop}%
\bibitem [{\citenamefont {Lagendijk}\ and\ \citenamefont
  {Van~Tiggelen}(1996)}]{lagendijk1996resonant}%
  \BibitemOpen
  \bibfield  {author} {\bibinfo {author} {\bibfnamefont {A.}~\bibnamefont
  {Lagendijk}}\ and\ \bibinfo {author} {\bibfnamefont {B.~A.}\ \bibnamefont
  {Van~Tiggelen}},\ }\href@noop {} {\bibfield  {journal} {\bibinfo  {journal}
  {Phys. Rep.}\ }\textbf {\bibinfo {volume} {270}},\ \bibinfo {pages} {143}
  (\bibinfo {year} {1996})}\BibitemShut {NoStop}%
\bibitem [{\citenamefont {Sheng}(2006)}]{sheng2006introduction}%
  \BibitemOpen
  \bibfield  {author} {\bibinfo {author} {\bibfnamefont {P.}~\bibnamefont
  {Sheng}},\ }\href@noop {} {\emph {\bibinfo {title} {Introduction to Wave
  Scattering, Localization and Mesoscopic Phenomena}}}\ (\bibinfo  {publisher}
  {Springer Science \& Business Media},\ \bibinfo {year} {2006})\BibitemShut
  {NoStop}%
\bibitem [{\citenamefont {Kumar}\ and\ \citenamefont
  {Tien}(1990)}]{kumar1990dependent}%
  \BibitemOpen
  \bibfield  {author} {\bibinfo {author} {\bibfnamefont {S.}~\bibnamefont
  {Kumar}}\ and\ \bibinfo {author} {\bibfnamefont {C.}~\bibnamefont {Tien}},\
  }\href@noop {} {\bibfield  {journal} {\bibinfo  {journal} {Journal of heat
  transfer}\ }\textbf {\bibinfo {volume} {112}},\ \bibinfo {pages} {178}
  (\bibinfo {year} {1990})}\BibitemShut {NoStop}%
\bibitem [{\citenamefont {Lee}(1992)}]{leeJTHT1992}%
  \BibitemOpen
  \bibfield  {author} {\bibinfo {author} {\bibfnamefont {S.-C.}\ \bibnamefont
  {Lee}},\ }\href@noop {} {\bibfield  {journal} {\bibinfo  {journal} {Journal
  of thermophysics and heat transfer}\ }\textbf {\bibinfo {volume} {6}},\
  \bibinfo {pages} {589} (\bibinfo {year} {1992})}\BibitemShut {NoStop}%
\bibitem [{\citenamefont {Ivezić}\ and\ \citenamefont
  {Mengüç}(1996)}]{ivezicIJHMT1996}%
  \BibitemOpen
  \bibfield  {author} {\bibinfo {author} {\bibfnamefont {Z.}~\bibnamefont
  {Ivezić}}\ and\ \bibinfo {author} {\bibfnamefont {M.~P.}\ \bibnamefont
  {Mengüç}},\ }\href {\doibase https://doi.org/10.1016/0017-9310(95)00142-5}
  {\bibfield  {journal} {\bibinfo  {journal} {International Journal of Heat and
  Mass Transfer}\ }\textbf {\bibinfo {volume} {39}},\ \bibinfo {pages} {811 }
  (\bibinfo {year} {1996})}\BibitemShut {NoStop}%
\bibitem [{\citenamefont {Durant}\ \emph {et~al.}(2007)\citenamefont {Durant},
  \citenamefont {Calvo-Perez}, \citenamefont {Vukadinovic},\ and\ \citenamefont
  {Greffet}}]{durantJOSAA2007}%
  \BibitemOpen
  \bibfield  {author} {\bibinfo {author} {\bibfnamefont {S.}~\bibnamefont
  {Durant}}, \bibinfo {author} {\bibfnamefont {O.}~\bibnamefont {Calvo-Perez}},
  \bibinfo {author} {\bibfnamefont {N.}~\bibnamefont {Vukadinovic}}, \ and\
  \bibinfo {author} {\bibfnamefont {J.-J.}\ \bibnamefont {Greffet}},\ }\href
  {\doibase 10.1364/JOSAA.24.002953} {\bibfield  {journal} {\bibinfo  {journal}
  {J. Opt. Soc. Am. A}\ }\textbf {\bibinfo {volume} {24}},\ \bibinfo {pages}
  {2953} (\bibinfo {year} {2007})}\BibitemShut {NoStop}%
\bibitem [{\citenamefont {Nguyen}\ \emph {et~al.}(2013)\citenamefont {Nguyen},
  \citenamefont {Faber}, \citenamefont {van~der Pol}, \citenamefont {van
  Leeuwen},\ and\ \citenamefont {Kalkman}}]{nguyenOE2013}%
  \BibitemOpen
  \bibfield  {author} {\bibinfo {author} {\bibfnamefont {V.~D.}\ \bibnamefont
  {Nguyen}}, \bibinfo {author} {\bibfnamefont {D.~J.}\ \bibnamefont {Faber}},
  \bibinfo {author} {\bibfnamefont {E.}~\bibnamefont {van~der Pol}}, \bibinfo
  {author} {\bibfnamefont {T.~G.}\ \bibnamefont {van Leeuwen}}, \ and\ \bibinfo
  {author} {\bibfnamefont {J.}~\bibnamefont {Kalkman}},\ }\href {\doibase
  10.1364/OE.21.029145} {\bibfield  {journal} {\bibinfo  {journal} {Opt.
  Express}\ }\textbf {\bibinfo {volume} {21}},\ \bibinfo {pages} {29145}
  (\bibinfo {year} {2013})}\BibitemShut {NoStop}%
\bibitem [{\citenamefont {Wang}\ and\ \citenamefont
  {Zhao}(2015)}]{wangIJHMT2015}%
  \BibitemOpen
  \bibfield  {author} {\bibinfo {author} {\bibfnamefont {B.~X.}\ \bibnamefont
  {Wang}}\ and\ \bibinfo {author} {\bibfnamefont {C.~Y.}\ \bibnamefont
  {Zhao}},\ }\href {\doibase 10.1016/j.ijheatmasstransfer.2015.06.017}
  {\bibfield  {journal} {\bibinfo  {journal} {International Journal of Heat and
  Mass Transfer}\ }\textbf {\bibinfo {volume} {89}},\ \bibinfo {pages} {920}
  (\bibinfo {year} {2015})}\BibitemShut {NoStop}%
\bibitem [{\citenamefont {Ma}\ \emph {et~al.}(2017)\citenamefont {Ma},
  \citenamefont {Tan}, \citenamefont {Zhao}, \citenamefont {Wang},\ and\
  \citenamefont {Wang}}]{maJQSRT2017}%
  \BibitemOpen
  \bibfield  {author} {\bibinfo {author} {\bibfnamefont {L.}~\bibnamefont
  {Ma}}, \bibinfo {author} {\bibfnamefont {J.}~\bibnamefont {Tan}}, \bibinfo
  {author} {\bibfnamefont {J.}~\bibnamefont {Zhao}}, \bibinfo {author}
  {\bibfnamefont {F.}~\bibnamefont {Wang}}, \ and\ \bibinfo {author}
  {\bibfnamefont {C.}~\bibnamefont {Wang}},\ }\href@noop {} {\bibfield
  {journal} {\bibinfo  {journal} {Journal of Quantitative Spectroscopy and
  Radiative Transfer}\ }\textbf {\bibinfo {volume} {187}},\ \bibinfo {pages}
  {255} (\bibinfo {year} {2017})}\BibitemShut {NoStop}%
\bibitem [{\citenamefont {Yamada}\ \emph {et~al.}(1986)\citenamefont {Yamada},
  \citenamefont {Cartigny},\ and\ \citenamefont {Tien}}]{yamadaJHT1986}%
  \BibitemOpen
  \bibfield  {author} {\bibinfo {author} {\bibfnamefont {Y.}~\bibnamefont
  {Yamada}}, \bibinfo {author} {\bibfnamefont {J.~D.}\ \bibnamefont
  {Cartigny}}, \ and\ \bibinfo {author} {\bibfnamefont {C.~L.}\ \bibnamefont
  {Tien}},\ }\href {\doibase 10.1115/1.3246980} {\bibfield  {journal} {\bibinfo
   {journal} {Journal of Heat Transfer}\ }\textbf {\bibinfo {volume} {108}}
  (\bibinfo {year} {1986}),\ 10.1115/1.3246980}\BibitemShut {NoStop}%
\bibitem [{\citenamefont {Aernouts}\ \emph {et~al.}(2014)\citenamefont
  {Aernouts}, \citenamefont {Beers}, \citenamefont {Watt{\'{e}}},\ and\
  \citenamefont {Lammertyn}}]{aernoutsOE2014}%
  \BibitemOpen
  \bibfield  {author} {\bibinfo {author} {\bibfnamefont {B.}~\bibnamefont
  {Aernouts}}, \bibinfo {author} {\bibfnamefont {R.~V.}\ \bibnamefont {Beers}},
  \bibinfo {author} {\bibfnamefont {R.}~\bibnamefont {Watt{\'{e}}}}, \ and\
  \bibinfo {author} {\bibfnamefont {J.}~\bibnamefont {Lammertyn}},\ }\href
  {\doibase 10.1364/OE.22.006086} {\bibfield  {journal} {\bibinfo  {journal}
  {Optics Express}\ }\textbf {\bibinfo {volume} {22}},\ \bibinfo {pages} {993}
  (\bibinfo {year} {2014})}\BibitemShut {NoStop}%
\bibitem [{\citenamefont {van Tiggelen}\ \emph {et~al.}(1990)\citenamefont {van
  Tiggelen}, \citenamefont {Lagendijk},\ and\ \citenamefont
  {Tip}}]{Vantiggelen1990JPCM}%
  \BibitemOpen
  \bibfield  {author} {\bibinfo {author} {\bibfnamefont {B.~A.}\ \bibnamefont
  {van Tiggelen}}, \bibinfo {author} {\bibfnamefont {A.}~\bibnamefont
  {Lagendijk}}, \ and\ \bibinfo {author} {\bibfnamefont {A.}~\bibnamefont
  {Tip}},\ }\href {http://stacks.iop.org/0953-8984/2/i=37/a=010} {\bibfield
  {journal} {\bibinfo  {journal} {J. Phys.: Cond. Mat.}\ }\textbf {\bibinfo
  {volume} {2}},\ \bibinfo {pages} {7653} (\bibinfo {year} {1990})}\BibitemShut
  {NoStop}%
\bibitem [{\citenamefont {Cherroret}\ \emph {et~al.}(2016)\citenamefont
  {Cherroret}, \citenamefont {Delande},\ and\ \citenamefont {van
  Tiggelen}}]{Cherroret2016}%
  \BibitemOpen
  \bibfield  {author} {\bibinfo {author} {\bibfnamefont {N.}~\bibnamefont
  {Cherroret}}, \bibinfo {author} {\bibfnamefont {D.}~\bibnamefont {Delande}},
  \ and\ \bibinfo {author} {\bibfnamefont {B.~A.}\ \bibnamefont {van
  Tiggelen}},\ }\href {\doibase 10.1103/PhysRevA.94.012702} {\bibfield
  {journal} {\bibinfo  {journal} {Phys. Rev. A}\ }\textbf {\bibinfo {volume}
  {94}},\ \bibinfo {pages} {012702} (\bibinfo {year} {2016})}\BibitemShut
  {NoStop}%
\bibitem [{\citenamefont {Wiersma}\ \emph {et~al.}(1997)\citenamefont
  {Wiersma}, \citenamefont {Bartolini}, \citenamefont {Lagendijk},\ and\
  \citenamefont {Righini}}]{wiersma1997localization}%
  \BibitemOpen
  \bibfield  {author} {\bibinfo {author} {\bibfnamefont {D.~S.}\ \bibnamefont
  {Wiersma}}, \bibinfo {author} {\bibfnamefont {P.}~\bibnamefont {Bartolini}},
  \bibinfo {author} {\bibfnamefont {A.}~\bibnamefont {Lagendijk}}, \ and\
  \bibinfo {author} {\bibfnamefont {R.}~\bibnamefont {Righini}},\ }\href@noop
  {} {\bibfield  {journal} {\bibinfo  {journal} {Nature (London)}\ }\textbf
  {\bibinfo {volume} {390}},\ \bibinfo {pages} {671} (\bibinfo {year}
  {1997})}\BibitemShut {NoStop}%
\bibitem [{\citenamefont {Ma}\ \emph {et~al.}(1990)\citenamefont {Ma},
  \citenamefont {Varadan},\ and\ \citenamefont {Varadan}}]{ma1990enhanced}%
  \BibitemOpen
  \bibfield  {author} {\bibinfo {author} {\bibfnamefont {Y.}~\bibnamefont
  {Ma}}, \bibinfo {author} {\bibfnamefont {V.}~\bibnamefont {Varadan}}, \ and\
  \bibinfo {author} {\bibfnamefont {V.}~\bibnamefont {Varadan}},\ }\href@noop
  {} {\bibfield  {journal} {\bibinfo  {journal} {Journal of heat transfer}\
  }\textbf {\bibinfo {volume} {112}},\ \bibinfo {pages} {402} (\bibinfo {year}
  {1990})}\BibitemShut {NoStop}%
\bibitem [{\citenamefont {Prasher}(2007)}]{prasherJAP2007}%
  \BibitemOpen
  \bibfield  {author} {\bibinfo {author} {\bibfnamefont {R.}~\bibnamefont
  {Prasher}},\ }\href {\doibase 10.1063/1.2794703} {\bibfield  {journal}
  {\bibinfo  {journal} {Journal of Applied Physics}\ }\textbf {\bibinfo
  {volume} {102}},\ \bibinfo {pages} {074316} (\bibinfo {year}
  {2007})}\BibitemShut {NoStop}%
\bibitem [{\citenamefont {Wei}\ \emph {et~al.}(2012)\citenamefont {Wei},
  \citenamefont {Fedorov}, \citenamefont {Luo},\ and\ \citenamefont
  {Ni}}]{weiAO2012}%
  \BibitemOpen
  \bibfield  {author} {\bibinfo {author} {\bibfnamefont {W.}~\bibnamefont
  {Wei}}, \bibinfo {author} {\bibfnamefont {A.~G.}\ \bibnamefont {Fedorov}},
  \bibinfo {author} {\bibfnamefont {Z.}~\bibnamefont {Luo}}, \ and\ \bibinfo
  {author} {\bibfnamefont {M.}~\bibnamefont {Ni}},\ }\href {\doibase
  10.1364/AO.51.006159} {\bibfield  {journal} {\bibinfo  {journal} {Appl.
  Opt.}\ }\textbf {\bibinfo {volume} {51}},\ \bibinfo {pages} {6159} (\bibinfo
  {year} {2012})}\BibitemShut {NoStop}%
\bibitem [{\citenamefont {Taylor}\ \emph {et~al.}(2013)\citenamefont {Taylor},
  \citenamefont {Coulombe}, \citenamefont {Otanicar}, \citenamefont {Phelan},
  \citenamefont {Gunawan}, \citenamefont {Lv}, \citenamefont {Rosengarten},
  \citenamefont {Prasher},\ and\ \citenamefont {Tyagi}}]{taylorJAP2013}%
  \BibitemOpen
  \bibfield  {author} {\bibinfo {author} {\bibfnamefont {R.}~\bibnamefont
  {Taylor}}, \bibinfo {author} {\bibfnamefont {S.}~\bibnamefont {Coulombe}},
  \bibinfo {author} {\bibfnamefont {T.}~\bibnamefont {Otanicar}}, \bibinfo
  {author} {\bibfnamefont {P.}~\bibnamefont {Phelan}}, \bibinfo {author}
  {\bibfnamefont {A.}~\bibnamefont {Gunawan}}, \bibinfo {author} {\bibfnamefont
  {W.}~\bibnamefont {Lv}}, \bibinfo {author} {\bibfnamefont {G.}~\bibnamefont
  {Rosengarten}}, \bibinfo {author} {\bibfnamefont {R.}~\bibnamefont
  {Prasher}}, \ and\ \bibinfo {author} {\bibfnamefont {H.}~\bibnamefont
  {Tyagi}},\ }\href {\doibase 10.1063/1.4754271} {\bibfield  {journal}
  {\bibinfo  {journal} {Journal of Applied Physics}\ }\textbf {\bibinfo
  {volume} {113}},\ \bibinfo {pages} {011301} (\bibinfo {year} {2013})},\
  \Eprint {http://arxiv.org/abs/https://doi.org/10.1063/1.4754271}
  {https://doi.org/10.1063/1.4754271} \BibitemShut {NoStop}%
\bibitem [{\citenamefont {Said}\ \emph {et~al.}(2013)\citenamefont {Said},
  \citenamefont {Sajid}, \citenamefont {Saidur}, \citenamefont
  {Kamalisarvestani},\ and\ \citenamefont {Rahim}}]{saidICHMT2013}%
  \BibitemOpen
  \bibfield  {author} {\bibinfo {author} {\bibfnamefont {Z.}~\bibnamefont
  {Said}}, \bibinfo {author} {\bibfnamefont {M.}~\bibnamefont {Sajid}},
  \bibinfo {author} {\bibfnamefont {R.}~\bibnamefont {Saidur}}, \bibinfo
  {author} {\bibfnamefont {M.}~\bibnamefont {Kamalisarvestani}}, \ and\
  \bibinfo {author} {\bibfnamefont {N.}~\bibnamefont {Rahim}},\ }\href
  {\doibase https://doi.org/10.1016/j.icheatmasstransfer.2013.05.013}
  {\bibfield  {journal} {\bibinfo  {journal} {International Communications in
  Heat and Mass Transfer}\ }\textbf {\bibinfo {volume} {46}},\ \bibinfo {pages}
  {74 } (\bibinfo {year} {2013})}\BibitemShut {NoStop}%
\bibitem [{\citenamefont {Xuan}\ \emph {et~al.}(2014)\citenamefont {Xuan},
  \citenamefont {Duan},\ and\ \citenamefont {Li}}]{xuanRSCA2014}%
  \BibitemOpen
  \bibfield  {author} {\bibinfo {author} {\bibfnamefont {Y.}~\bibnamefont
  {Xuan}}, \bibinfo {author} {\bibfnamefont {H.}~\bibnamefont {Duan}}, \ and\
  \bibinfo {author} {\bibfnamefont {Q.}~\bibnamefont {Li}},\ }\href {\doibase
  10.1039/C4RA00630E} {\bibfield  {journal} {\bibinfo  {journal} {RSC Adv.}\
  }\textbf {\bibinfo {volume} {4}},\ \bibinfo {pages} {16206} (\bibinfo {year}
  {2014})}\BibitemShut {NoStop}%
\bibitem [{\citenamefont {Hogan}\ \emph {et~al.}(2014)\citenamefont {Hogan},
  \citenamefont {Urban}, \citenamefont {Ayala-Orozco}, \citenamefont
  {Pimpinelli}, \citenamefont {Nordlander},\ and\ \citenamefont
  {Halas}}]{hoganNL2014}%
  \BibitemOpen
  \bibfield  {author} {\bibinfo {author} {\bibfnamefont {N.~J.}\ \bibnamefont
  {Hogan}}, \bibinfo {author} {\bibfnamefont {A.~S.}\ \bibnamefont {Urban}},
  \bibinfo {author} {\bibfnamefont {C.}~\bibnamefont {Ayala-Orozco}}, \bibinfo
  {author} {\bibfnamefont {A.}~\bibnamefont {Pimpinelli}}, \bibinfo {author}
  {\bibfnamefont {P.}~\bibnamefont {Nordlander}}, \ and\ \bibinfo {author}
  {\bibfnamefont {N.~J.}\ \bibnamefont {Halas}},\ }\href {\doibase
  10.1021/nl5016975} {\bibfield  {journal} {\bibinfo  {journal} {Nano Letters}\
  }\textbf {\bibinfo {volume} {14}},\ \bibinfo {pages} {4640} (\bibinfo {year}
  {2014})},\ \bibinfo {note} {pMID: 24960442},\ \Eprint
  {http://arxiv.org/abs/http://dx.doi.org/10.1021/nl5016975}
  {http://dx.doi.org/10.1021/nl5016975} \BibitemShut {NoStop}%
\bibitem [{\citenamefont {Liu}\ and\ \citenamefont
  {Xuan}(2017)}]{liuNanoscale2017}%
  \BibitemOpen
  \bibfield  {author} {\bibinfo {author} {\bibfnamefont {X.}~\bibnamefont
  {Liu}}\ and\ \bibinfo {author} {\bibfnamefont {Y.}~\bibnamefont {Xuan}},\
  }\href {\doibase 10.1039/C7NR03912C} {\bibfield  {journal} {\bibinfo
  {journal} {Nanoscale}\ }\textbf {\bibinfo {volume} {9}},\ \bibinfo {pages}
  {14854} (\bibinfo {year} {2017})}\BibitemShut {NoStop}%
\bibitem [{\citenamefont {Gao}\ \emph {et~al.}(2017)\citenamefont {Gao},
  \citenamefont {Zhao},\ and\ \citenamefont {Wang}}]{gaoJAP2017}%
  \BibitemOpen
  \bibfield  {author} {\bibinfo {author} {\bibfnamefont {J.~D.}\ \bibnamefont
  {Gao}}, \bibinfo {author} {\bibfnamefont {C.~Y.}\ \bibnamefont {Zhao}}, \
  and\ \bibinfo {author} {\bibfnamefont {B.~X.}\ \bibnamefont {Wang}},\ }\href
  {\doibase 10.1063/1.4978418} {\bibfield  {journal} {\bibinfo  {journal}
  {Journal of Applied Physics}\ }\textbf {\bibinfo {volume} {121}},\ \bibinfo
  {pages} {113105} (\bibinfo {year} {2017})},\ \Eprint
  {http://arxiv.org/abs/https://doi.org/10.1063/1.4978418}
  {https://doi.org/10.1063/1.4978418} \BibitemShut {NoStop}%
\bibitem [{\citenamefont {Psaltis}\ \emph {et~al.}(2006)\citenamefont
  {Psaltis}, \citenamefont {Quake},\ and\ \citenamefont
  {Yang}}]{psaltisNature2006}%
  \BibitemOpen
  \bibfield  {author} {\bibinfo {author} {\bibfnamefont {D.}~\bibnamefont
  {Psaltis}}, \bibinfo {author} {\bibfnamefont {S.~R.}\ \bibnamefont {Quake}},
  \ and\ \bibinfo {author} {\bibfnamefont {C.}~\bibnamefont {Yang}},\
  }\href@noop {} {\bibfield  {journal} {\bibinfo  {journal} {Nature}\ }\textbf
  {\bibinfo {volume} {442}},\ \bibinfo {pages} {381} (\bibinfo {year}
  {2006})}\BibitemShut {NoStop}%
\bibitem [{\citenamefont {Erickson}\ \emph {et~al.}(2011)\citenamefont
  {Erickson}, \citenamefont {Sinton},\ and\ \citenamefont
  {Psaltis}}]{ericksonNPhoton2011}%
  \BibitemOpen
  \bibfield  {author} {\bibinfo {author} {\bibfnamefont {D.}~\bibnamefont
  {Erickson}}, \bibinfo {author} {\bibfnamefont {D.}~\bibnamefont {Sinton}}, \
  and\ \bibinfo {author} {\bibfnamefont {D.}~\bibnamefont {Psaltis}},\
  }\href@noop {} {\bibfield  {journal} {\bibinfo  {journal} {Nature Photonics}\
  }\textbf {\bibinfo {volume} {5}},\ \bibinfo {pages} {583} (\bibinfo {year}
  {2011})}\BibitemShut {NoStop}%
\bibitem [{\citenamefont {Zielinski}\ \emph {et~al.}(2016)\citenamefont
  {Zielinski}, \citenamefont {Choi}, \citenamefont {La~Grange}, \citenamefont
  {Modestino}, \citenamefont {Hashemi}, \citenamefont {Pu}, \citenamefont
  {Birkhold}, \citenamefont {Hubbell},\ and\ \citenamefont
  {Psaltis}}]{zielinskiNL2016}%
  \BibitemOpen
  \bibfield  {author} {\bibinfo {author} {\bibfnamefont {M.~S.}\ \bibnamefont
  {Zielinski}}, \bibinfo {author} {\bibfnamefont {J.-W.}\ \bibnamefont {Choi}},
  \bibinfo {author} {\bibfnamefont {T.}~\bibnamefont {La~Grange}}, \bibinfo
  {author} {\bibfnamefont {M.}~\bibnamefont {Modestino}}, \bibinfo {author}
  {\bibfnamefont {S.~M.~H.}\ \bibnamefont {Hashemi}}, \bibinfo {author}
  {\bibfnamefont {Y.}~\bibnamefont {Pu}}, \bibinfo {author} {\bibfnamefont
  {S.}~\bibnamefont {Birkhold}}, \bibinfo {author} {\bibfnamefont {J.~A.}\
  \bibnamefont {Hubbell}}, \ and\ \bibinfo {author} {\bibfnamefont
  {D.}~\bibnamefont {Psaltis}},\ }\href {\doibase 10.1021/acs.nanolett.5b03901}
  {\bibfield  {journal} {\bibinfo  {journal} {Nano letters}\ }\textbf {\bibinfo
  {volume} {16}},\ \bibinfo {pages} {2159} (\bibinfo {year}
  {2016})}\BibitemShut {NoStop}%
\bibitem [{\citenamefont {Bohren}\ and\ \citenamefont
  {Huffman}(2008)}]{bohrenandhuffman}%
  \BibitemOpen
  \bibfield  {author} {\bibinfo {author} {\bibfnamefont {C.~F.}\ \bibnamefont
  {Bohren}}\ and\ \bibinfo {author} {\bibfnamefont {D.~R.}\ \bibnamefont
  {Huffman}},\ }\href@noop {} {\emph {\bibinfo {title} {Absorption and
  scattering of light by small particles}}}\ (\bibinfo  {publisher} {John Wiley
  \& Sons},\ \bibinfo {year} {2008})\BibitemShut {NoStop}%
\bibitem [{\citenamefont {Markel}(1993)}]{markel1993}%
  \BibitemOpen
  \bibfield  {author} {\bibinfo {author} {\bibfnamefont {V.}~\bibnamefont
  {Markel}},\ }\href {\doibase 10.1080/09500349314552291} {\bibfield  {journal}
  {\bibinfo  {journal} {Journal of Modern Optics}\ }\textbf {\bibinfo {volume}
  {40}},\ \bibinfo {pages} {2281} (\bibinfo {year} {1993})},\ \Eprint
  {http://arxiv.org/abs/https://doi.org/10.1080/09500349314552291}
  {https://doi.org/10.1080/09500349314552291} \BibitemShut {NoStop}%
\bibitem [{\citenamefont {Leung~Tsang}\ and\ \citenamefont
  {Ding}(2000)}]{tsang2000scattering1}%
  \BibitemOpen
  \bibfield  {author} {\bibinfo {author} {\bibfnamefont {J.~A.~K.}\
  \bibnamefont {Leung~Tsang}}\ and\ \bibinfo {author} {\bibfnamefont {K.~H.}\
  \bibnamefont {Ding}},\ }\href@noop {} {\emph {\bibinfo {title} {Scattering of
  Electromagnetic Waves, Theories and Applications, vol. 1}}}\ (\bibinfo
  {publisher} {New York: Wiley},\ \bibinfo {year} {2000})\BibitemShut {NoStop}%
\bibitem [{\citenamefont {Schuurmans}\ \emph {et~al.}(1999)\citenamefont
  {Schuurmans}, , \citenamefont {Vanmaekelbergh}, \citenamefont {Lagemaat},\
  and\ \citenamefont {Lagendijk}}]{schuurmansScience1999}%
  \BibitemOpen
  \bibfield  {author} {\bibinfo {author} {\bibfnamefont {F.~J.~P.}\
  \bibnamefont {Schuurmans}}, , \bibinfo {author} {\bibfnamefont
  {D.}~\bibnamefont {Vanmaekelbergh}}, \bibinfo {author} {\bibfnamefont
  {J.~v.~d.}\ \bibnamefont {Lagemaat}}, \ and\ \bibinfo {author} {\bibfnamefont
  {A.}~\bibnamefont {Lagendijk}},\ }\href {\doibase
  10.1126/science.284.5411.141} {\bibfield  {journal} {\bibinfo  {journal}
  {Science}\ }\textbf {\bibinfo {volume} {284}},\ \bibinfo {pages} {141}
  (\bibinfo {year} {1999})},\ \Eprint
  {http://arxiv.org/abs/http://science.sciencemag.org/content/284/5411/141.full.pdf}
  {http://science.sciencemag.org/content/284/5411/141.full.pdf} \BibitemShut
  {NoStop}%
\bibitem [{\citenamefont {Wang}\ \emph {et~al.}(1995)\citenamefont {Wang},
  \citenamefont {Jacques},\ and\ \citenamefont {Zheng}}]{wang1995mcml}%
  \BibitemOpen
  \bibfield  {author} {\bibinfo {author} {\bibfnamefont {L.}~\bibnamefont
  {Wang}}, \bibinfo {author} {\bibfnamefont {S.~L.}\ \bibnamefont {Jacques}}, \
  and\ \bibinfo {author} {\bibfnamefont {L.}~\bibnamefont {Zheng}},\
  }\href@noop {} {\bibfield  {journal} {\bibinfo  {journal} {Computer methods
  and programs in biomedicine}\ }\textbf {\bibinfo {volume} {47}},\ \bibinfo
  {pages} {131} (\bibinfo {year} {1995})}\BibitemShut {NoStop}%
\bibitem [{\citenamefont {Lax}(1952)}]{laxPR1952}%
  \BibitemOpen
  \bibfield  {author} {\bibinfo {author} {\bibfnamefont {M.}~\bibnamefont
  {Lax}},\ }\href {\doibase 10.1103/PhysRev.85.621} {\bibfield  {journal}
  {\bibinfo  {journal} {Physical Review}\ }\textbf {\bibinfo {volume} {85}},\
  \bibinfo {pages} {621} (\bibinfo {year} {1952})}\BibitemShut {NoStop}%
\bibitem [{\citenamefont {Roth}(1974)}]{rothPRB1974}%
  \BibitemOpen
  \bibfield  {author} {\bibinfo {author} {\bibfnamefont {L.~M.}\ \bibnamefont
  {Roth}},\ }\href {\doibase 10.1103/PhysRevB.9.2476} {\bibfield  {journal}
  {\bibinfo  {journal} {Phys. Rev. B}\ }\textbf {\bibinfo {volume} {9}},\
  \bibinfo {pages} {2476} (\bibinfo {year} {1974})}\BibitemShut {NoStop}%
\bibitem [{\citenamefont {Davis}\ and\ \citenamefont
  {Schwartz}(1985)}]{davisPRB1985}%
  \BibitemOpen
  \bibfield  {author} {\bibinfo {author} {\bibfnamefont {V.~A.}\ \bibnamefont
  {Davis}}\ and\ \bibinfo {author} {\bibfnamefont {L.}~\bibnamefont
  {Schwartz}},\ }\href {\doibase 10.1103/PhysRevB.31.5155} {\bibfield
  {journal} {\bibinfo  {journal} {Phys. Rev. B}\ }\textbf {\bibinfo {volume}
  {31}},\ \bibinfo {pages} {5155} (\bibinfo {year} {1985})}\BibitemShut
  {NoStop}%
\bibitem [{\citenamefont {Liebsch}\ and\ \citenamefont
  {Persson}(1983)}]{liebschJPC1983}%
  \BibitemOpen
  \bibfield  {author} {\bibinfo {author} {\bibfnamefont {A.}~\bibnamefont
  {Liebsch}}\ and\ \bibinfo {author} {\bibfnamefont {B.~N.~J.}\ \bibnamefont
  {Persson}},\ }\href {http://stacks.iop.org/0022-3719/16/i=27/a=019}
  {\bibfield  {journal} {\bibinfo  {journal} {Journal of Physics C: Solid State
  Physics}\ }\textbf {\bibinfo {volume} {16}},\ \bibinfo {pages} {5375}
  (\bibinfo {year} {1983})}\BibitemShut {NoStop}%
\bibitem [{\citenamefont {Liebsch}\ and\ \citenamefont
  {Gonz\'alez}(1984)}]{liebschPRB1984}%
  \BibitemOpen
  \bibfield  {author} {\bibinfo {author} {\bibfnamefont {A.}~\bibnamefont
  {Liebsch}}\ and\ \bibinfo {author} {\bibfnamefont {P.~V.~n.}\ \bibnamefont
  {Gonz\'alez}},\ }\href {\doibase 10.1103/PhysRevB.29.6907} {\bibfield
  {journal} {\bibinfo  {journal} {Phys. Rev. B}\ }\textbf {\bibinfo {volume}
  {29}},\ \bibinfo {pages} {6907} (\bibinfo {year} {1984})}\BibitemShut
  {NoStop}%
\bibitem [{\citenamefont {Felderhof}\ \emph {et~al.}(1983)\citenamefont
  {Felderhof}, \citenamefont {Ford},\ and\ \citenamefont
  {Cohen}}]{Felderhof1983}%
  \BibitemOpen
  \bibfield  {author} {\bibinfo {author} {\bibfnamefont {B.~U.}\ \bibnamefont
  {Felderhof}}, \bibinfo {author} {\bibfnamefont {G.~W.}\ \bibnamefont {Ford}},
  \ and\ \bibinfo {author} {\bibfnamefont {E.~G.~D.}\ \bibnamefont {Cohen}},\
  }\href {\doibase 10.1007/BF01009796} {\bibfield  {journal} {\bibinfo
  {journal} {Journal of Statistical Physics}\ }\textbf {\bibinfo {volume}
  {33}},\ \bibinfo {pages} {241} (\bibinfo {year} {1983})}\BibitemShut
  {NoStop}%
\bibitem [{\citenamefont {Torquato}(1984)}]{torquatoJCP1984}%
  \BibitemOpen
  \bibfield  {author} {\bibinfo {author} {\bibfnamefont {S.}~\bibnamefont
  {Torquato}},\ }\href {\doibase 10.1063/1.447497} {\bibfield  {journal}
  {\bibinfo  {journal} {The Journal of Chemical Physics}\ }\textbf {\bibinfo
  {volume} {81}},\ \bibinfo {pages} {5079} (\bibinfo {year}
  {1984})}\BibitemShut {NoStop}%
\bibitem [{\citenamefont {Barrera}\ \emph {et~al.}(1988)\citenamefont
  {Barrera}, \citenamefont {Monsivais},\ and\ \citenamefont
  {Moch\'an}}]{barreraPRB1988}%
  \BibitemOpen
  \bibfield  {author} {\bibinfo {author} {\bibfnamefont {R.~G.}\ \bibnamefont
  {Barrera}}, \bibinfo {author} {\bibfnamefont {G.}~\bibnamefont {Monsivais}},
  \ and\ \bibinfo {author} {\bibfnamefont {W.~L.}\ \bibnamefont {Moch\'an}},\
  }\href {\doibase 10.1103/PhysRevB.38.5371} {\bibfield  {journal} {\bibinfo
  {journal} {Phys. Rev. B}\ }\textbf {\bibinfo {volume} {38}},\ \bibinfo
  {pages} {5371} (\bibinfo {year} {1988})}\BibitemShut {NoStop}%
\bibitem [{\citenamefont {Barrera}\ \emph {et~al.}(1989)\citenamefont
  {Barrera}, \citenamefont {Monsiv\'ais}, \citenamefont {Moch\'an},\ and\
  \citenamefont {Anda}}]{barreraPRB1989}%
  \BibitemOpen
  \bibfield  {author} {\bibinfo {author} {\bibfnamefont {R.~G.}\ \bibnamefont
  {Barrera}}, \bibinfo {author} {\bibfnamefont {G.}~\bibnamefont
  {Monsiv\'ais}}, \bibinfo {author} {\bibfnamefont {W.~L.}\ \bibnamefont
  {Moch\'an}}, \ and\ \bibinfo {author} {\bibfnamefont {E.}~\bibnamefont
  {Anda}},\ }\href {\doibase 10.1103/PhysRevB.39.9998} {\bibfield  {journal}
  {\bibinfo  {journal} {Phys. Rev. B}\ }\textbf {\bibinfo {volume} {39}},\
  \bibinfo {pages} {9998} (\bibinfo {year} {1989})}\BibitemShut {NoStop}%
\bibitem [{\citenamefont {Claro}\ and\ \citenamefont
  {Rojas}(1991)}]{Claro1991}%
  \BibitemOpen
  \bibfield  {author} {\bibinfo {author} {\bibfnamefont {F.}~\bibnamefont
  {Claro}}\ and\ \bibinfo {author} {\bibfnamefont {R.}~\bibnamefont {Rojas}},\
  }\href {\doibase 10.1103/PhysRevB.43.6369} {\bibfield  {journal} {\bibinfo
  {journal} {Phys. Rev. B}\ }\textbf {\bibinfo {volume} {43}},\ \bibinfo
  {pages} {6369} (\bibinfo {year} {1991})}\BibitemShut {NoStop}%
\bibitem [{\citenamefont {Vasilevskiy}\ and\ \citenamefont
  {Anda}(1996)}]{vasilevskiyPRB1996}%
  \BibitemOpen
  \bibfield  {author} {\bibinfo {author} {\bibfnamefont {M.~I.}\ \bibnamefont
  {Vasilevskiy}}\ and\ \bibinfo {author} {\bibfnamefont {E.~V.}\ \bibnamefont
  {Anda}},\ }\href {\doibase 10.1103/PhysRevB.54.5844} {\bibfield  {journal}
  {\bibinfo  {journal} {Phys. Rev. B}\ }\textbf {\bibinfo {volume} {54}},\
  \bibinfo {pages} {5844} (\bibinfo {year} {1996})}\BibitemShut {NoStop}%
\bibitem [{\citenamefont {Ersfeld}\ and\ \citenamefont
  {Felderhof}(1998)}]{felderhofPRE1998}%
  \BibitemOpen
  \bibfield  {author} {\bibinfo {author} {\bibfnamefont {B.}~\bibnamefont
  {Ersfeld}}\ and\ \bibinfo {author} {\bibfnamefont {B.~U.}\ \bibnamefont
  {Felderhof}},\ }\href {\doibase 10.1103/PhysRevE.57.1118} {\bibfield
  {journal} {\bibinfo  {journal} {Phys. Rev. E}\ }\textbf {\bibinfo {volume}
  {57}},\ \bibinfo {pages} {1118} (\bibinfo {year} {1998})}\BibitemShut
  {NoStop}%
\bibitem [{\citenamefont {Lax}(1951)}]{laxRMP1951}%
  \BibitemOpen
  \bibfield  {author} {\bibinfo {author} {\bibfnamefont {M.}~\bibnamefont
  {Lax}},\ }\href {\doibase 10.1103/RevModPhys.23.287} {\bibfield  {journal}
  {\bibinfo  {journal} {Rev. Mod. Phys.}\ }\textbf {\bibinfo {volume} {23}},\
  \bibinfo {pages} {287} (\bibinfo {year} {1951})}\BibitemShut {NoStop}%
\bibitem [{\citenamefont {Foldy}(1945)}]{foldyPR1945}%
  \BibitemOpen
  \bibfield  {author} {\bibinfo {author} {\bibfnamefont {L.~L.}\ \bibnamefont
  {Foldy}},\ }\href {\doibase 10.1103/PhysRev.67.107} {\bibfield  {journal}
  {\bibinfo  {journal} {Phys. Rev.}\ }\textbf {\bibinfo {volume} {67}},\
  \bibinfo {pages} {107} (\bibinfo {year} {1945})}\BibitemShut {NoStop}%
\bibitem [{\citenamefont {Fraden}\ and\ \citenamefont
  {Maret}(1990)}]{fradenPRL1990}%
  \BibitemOpen
  \bibfield  {author} {\bibinfo {author} {\bibfnamefont {S.}~\bibnamefont
  {Fraden}}\ and\ \bibinfo {author} {\bibfnamefont {G.}~\bibnamefont {Maret}},\
  }\href {\doibase 10.1103/PhysRevLett.65.512} {\bibfield  {journal} {\bibinfo
  {journal} {Phys. Rev. Lett.}\ }\textbf {\bibinfo {volume} {65}},\ \bibinfo
  {pages} {512} (\bibinfo {year} {1990})}\BibitemShut {NoStop}%
\bibitem [{\citenamefont {Rojas-Ochoa}\ \emph {et~al.}(2004)\citenamefont
  {Rojas-Ochoa}, \citenamefont {Mendez-Alcaraz}, \citenamefont {S\'aenz},
  \citenamefont {Schurtenberger},\ and\ \citenamefont
  {Scheffold}}]{rojasochoaPRL2004}%
  \BibitemOpen
  \bibfield  {author} {\bibinfo {author} {\bibfnamefont {L.~F.}\ \bibnamefont
  {Rojas-Ochoa}}, \bibinfo {author} {\bibfnamefont {J.~M.}\ \bibnamefont
  {Mendez-Alcaraz}}, \bibinfo {author} {\bibfnamefont {J.~J.}\ \bibnamefont
  {S\'aenz}}, \bibinfo {author} {\bibfnamefont {P.}~\bibnamefont
  {Schurtenberger}}, \ and\ \bibinfo {author} {\bibfnamefont {F.}~\bibnamefont
  {Scheffold}},\ }\href {\doibase 10.1103/PhysRevLett.93.073903} {\bibfield
  {journal} {\bibinfo  {journal} {Phys. Rev. Lett.}\ }\textbf {\bibinfo
  {volume} {93}},\ \bibinfo {pages} {073903} (\bibinfo {year}
  {2004})}\BibitemShut {NoStop}%
\bibitem [{\citenamefont {Tsang}\ and\ \citenamefont
  {Kong}(1982)}]{tsangJAP1982}%
  \BibitemOpen
  \bibfield  {author} {\bibinfo {author} {\bibfnamefont {L.}~\bibnamefont
  {Tsang}}\ and\ \bibinfo {author} {\bibfnamefont {J.~A.}\ \bibnamefont
  {Kong}},\ }\href {\doibase 10.1063/1.331611} {\bibfield  {journal} {\bibinfo
  {journal} {Journal of Applied Physics}\ }\textbf {\bibinfo {volume} {53}},\
  \bibinfo {pages} {7162} (\bibinfo {year} {1982})}\BibitemShut {NoStop}%
\bibitem [{\citenamefont {Ma}\ \emph {et~al.}(1988)\citenamefont {Ma},
  \citenamefont {Varadan},\ and\ \citenamefont {Varadan}}]{maAO1988}%
  \BibitemOpen
  \bibfield  {author} {\bibinfo {author} {\bibfnamefont {Y.}~\bibnamefont
  {Ma}}, \bibinfo {author} {\bibfnamefont {V.~V.}\ \bibnamefont {Varadan}}, \
  and\ \bibinfo {author} {\bibfnamefont {V.~K.}\ \bibnamefont {Varadan}},\
  }\href {\doibase 10.1364/AO.27.002469} {\bibfield  {journal} {\bibinfo
  {journal} {Applied Optics}\ }\textbf {\bibinfo {volume} {27}},\ \bibinfo
  {pages} {2469} (\bibinfo {year} {1988})}\BibitemShut {NoStop}%
\bibitem [{\citenamefont {Barabanenkov}(1975)}]{barabanenkov1975multiple}%
  \BibitemOpen
  \bibfield  {author} {\bibinfo {author} {\bibfnamefont {Y.~N.}\ \bibnamefont
  {Barabanenkov}},\ }\href@noop {} {\bibfield  {journal} {\bibinfo  {journal}
  {Soviet Physics Uspekhi}\ }\textbf {\bibinfo {volume} {18}},\ \bibinfo
  {pages} {673} (\bibinfo {year} {1975})}\BibitemShut {NoStop}%
\bibitem [{\citenamefont {Baxter}(1968)}]{Baxter1968}%
  \BibitemOpen
  \bibfield  {author} {\bibinfo {author} {\bibfnamefont {R.~J.}\ \bibnamefont
  {Baxter}},\ }\href {\doibase http://dx.doi.org/10.1063/1.1670482} {\bibfield
  {journal} {\bibinfo  {journal} {J. Chem. Phys.}\ }\textbf {\bibinfo {volume}
  {49}},\ \bibinfo {pages} {2770} (\bibinfo {year} {1968})}\BibitemShut
  {NoStop}%
\bibitem [{\citenamefont {Tsang}\ \emph {et~al.}(2004)\citenamefont {Tsang},
  \citenamefont {Kong}, \citenamefont {Ding},\ and\ \citenamefont
  {Ao}}]{tsang2004scattering2}%
  \BibitemOpen
  \bibfield  {author} {\bibinfo {author} {\bibfnamefont {L.}~\bibnamefont
  {Tsang}}, \bibinfo {author} {\bibfnamefont {J.~A.}\ \bibnamefont {Kong}},
  \bibinfo {author} {\bibfnamefont {K.-H.}\ \bibnamefont {Ding}}, \ and\
  \bibinfo {author} {\bibfnamefont {C.~O.}\ \bibnamefont {Ao}},\ }\href@noop {}
  {\emph {\bibinfo {title} {Scattering of Electromagnetic Waves: Numerical
  Simulations}}}\ (\bibinfo  {publisher} {John Wiley \& Sons},\ \bibinfo {year}
  {2004})\BibitemShut {NoStop}%
\bibitem [{\citenamefont {Wen}\ \emph {et~al.}(1990)\citenamefont {Wen},
  \citenamefont {Tsang}, \citenamefont {Winebrenner},\ and\ \citenamefont
  {Ishimaru}}]{wen1990dense}%
  \BibitemOpen
  \bibfield  {author} {\bibinfo {author} {\bibfnamefont {B.}~\bibnamefont
  {Wen}}, \bibinfo {author} {\bibfnamefont {L.}~\bibnamefont {Tsang}}, \bibinfo
  {author} {\bibfnamefont {D.~P.}\ \bibnamefont {Winebrenner}}, \ and\ \bibinfo
  {author} {\bibfnamefont {A.}~\bibnamefont {Ishimaru}},\ }\href@noop {}
  {\bibfield  {journal} {\bibinfo  {journal} {IEEE Transactions on Geoscience
  and Remote Sensing}\ }\textbf {\bibinfo {volume} {28}},\ \bibinfo {pages}
  {46} (\bibinfo {year} {1990})}\BibitemShut {NoStop}%
\bibitem [{\citenamefont {Shore}\ and\ \citenamefont
  {Yaghjian}(2012)}]{shore2012complex}%
  \BibitemOpen
  \bibfield  {author} {\bibinfo {author} {\bibfnamefont {R.~A.}\ \bibnamefont
  {Shore}}\ and\ \bibinfo {author} {\bibfnamefont {A.~D.}\ \bibnamefont
  {Yaghjian}},\ }\href@noop {} {\bibfield  {journal} {\bibinfo  {journal}
  {Radio Science}\ }\textbf {\bibinfo {volume} {47}} (\bibinfo {year}
  {2012})}\BibitemShut {NoStop}%
\bibitem [{\citenamefont {West}\ \emph {et~al.}(1994)\citenamefont {West},
  \citenamefont {Gibbs}, \citenamefont {Tsang},\ and\ \citenamefont
  {Fung}}]{westJOSAA1994}%
  \BibitemOpen
  \bibfield  {author} {\bibinfo {author} {\bibfnamefont {R.}~\bibnamefont
  {West}}, \bibinfo {author} {\bibfnamefont {D.}~\bibnamefont {Gibbs}},
  \bibinfo {author} {\bibfnamefont {L.}~\bibnamefont {Tsang}}, \ and\ \bibinfo
  {author} {\bibfnamefont {A.~K.}\ \bibnamefont {Fung}},\ }\href {\doibase
  10.1364/JOSAA.11.001854} {\bibfield  {journal} {\bibinfo  {journal} {J. Opt.
  Soc. Am. A}\ }\textbf {\bibinfo {volume} {11}},\ \bibinfo {pages} {1854}
  (\bibinfo {year} {1994})}\BibitemShut {NoStop}%
\bibitem [{\citenamefont {Magkiriadou}\ \emph {et~al.}(2012)\citenamefont
  {Magkiriadou}, \citenamefont {Park}, \citenamefont {Kim},\ and\ \citenamefont
  {Manoharan}}]{magkiriadou2012disordered}%
  \BibitemOpen
  \bibfield  {author} {\bibinfo {author} {\bibfnamefont {S.}~\bibnamefont
  {Magkiriadou}}, \bibinfo {author} {\bibfnamefont {J.-G.}\ \bibnamefont
  {Park}}, \bibinfo {author} {\bibfnamefont {Y.-S.}\ \bibnamefont {Kim}}, \
  and\ \bibinfo {author} {\bibfnamefont {V.~N.}\ \bibnamefont {Manoharan}},\
  }\href@noop {} {\bibfield  {journal} {\bibinfo  {journal} {Optical Materials
  Express}\ }\textbf {\bibinfo {volume} {2}},\ \bibinfo {pages} {1343}
  (\bibinfo {year} {2012})}\BibitemShut {NoStop}%
\bibitem [{\citenamefont {van Tiggelen}\ \emph {et~al.}(2000)\citenamefont {van
  Tiggelen}, \citenamefont {Lagendijk},\ and\ \citenamefont
  {Wiersma}}]{tiggelenPRL2000}%
  \BibitemOpen
  \bibfield  {author} {\bibinfo {author} {\bibfnamefont {B.~A.}\ \bibnamefont
  {van Tiggelen}}, \bibinfo {author} {\bibfnamefont {A.}~\bibnamefont
  {Lagendijk}}, \ and\ \bibinfo {author} {\bibfnamefont {D.~S.}\ \bibnamefont
  {Wiersma}},\ }\href {\doibase 10.1103/PhysRevLett.84.4333} {\bibfield
  {journal} {\bibinfo  {journal} {Phys. Rev. Lett.}\ }\textbf {\bibinfo
  {volume} {84}},\ \bibinfo {pages} {4333} (\bibinfo {year}
  {2000})}\BibitemShut {NoStop}%
\bibitem [{\citenamefont {Yamilov}\ \emph {et~al.}(2014)\citenamefont
  {Yamilov}, \citenamefont {Sarma}, \citenamefont {Redding}, \citenamefont
  {Payne}, \citenamefont {Noh},\ and\ \citenamefont {Cao}}]{yamilovPRL2014}%
  \BibitemOpen
  \bibfield  {author} {\bibinfo {author} {\bibfnamefont {A.~G.}\ \bibnamefont
  {Yamilov}}, \bibinfo {author} {\bibfnamefont {R.}~\bibnamefont {Sarma}},
  \bibinfo {author} {\bibfnamefont {B.}~\bibnamefont {Redding}}, \bibinfo
  {author} {\bibfnamefont {B.}~\bibnamefont {Payne}}, \bibinfo {author}
  {\bibfnamefont {H.}~\bibnamefont {Noh}}, \ and\ \bibinfo {author}
  {\bibfnamefont {H.}~\bibnamefont {Cao}},\ }\href {\doibase
  10.1103/PhysRevLett.112.023904} {\bibfield  {journal} {\bibinfo  {journal}
  {Phys. Rev. Lett.}\ }\textbf {\bibinfo {volume} {112}},\ \bibinfo {pages}
  {023904} (\bibinfo {year} {2014})}\BibitemShut {NoStop}%
\end{thebibliography}%

\end{document}